%% file: main.tex
\documentclass[sigconf,10pt]{acmart}

\setcopyright{acmcopyright}
\copyrightyear{2022}
\acmYear{2022}
\acmDOI{XXXXXXX.XXXXXXX}

\usepackage{tikz}
\usepackage{xcolor}

\usepackage{multirow}
\usepackage{multicol}
\usepackage{xcolor}
\usepackage{pifont}
\usepackage{varioref}
\usepackage{eucal}
\usepackage{subfig}
\usepackage{rotating}
\usepackage{colortbl} 
\usepackage{amsmath,amsfonts}
\usepackage{enumerate}
\usepackage{tablefootnote}
\usepackage{threeparttable}
\usepackage{soul}
\usepackage{booktabs}
\usepackage{algorithm,algpseudocode}
\usepackage[font={footnotesize}]{caption}

\usepackage{enumitem,kantlipsum}
\usepackage{array}
\usepackage{tablefootnote}
\usepackage{threeparttable}
\usepackage[toc,title,page]{appendix}
\usepackage{textcomp}
\usepackage{bmpsize}
\def\BibTeX{{\rm B\kern-.05em{\sc i\kern-.025em b}\kern-.08em
    T\kern-.1667em\lower.7ex\hbox{E}\kern-.125emX}}

\newcolumntype{P}[1]{>{\centering\arraybackslash}p{#1}}

\newcommand{\spy}{\textit{EarSpy}}

\begin{document}

\sloppy

%\title{Is Two-Factor Authentication Secure in the Face of Malicious Terminals: A Comprehensive Study of 5 Prominent 2FA Deployments }
%\title{Is 2FA Secure in the Face of Compromised Terminals: A Study of Concurrent Login Attacks on Five Recent Deployments}
%\title{2FA vs.\ Compromised Terminals:\\ A Study of Concurrent Hidden Login Attacks on Five Prominent 2FA Deployments}
\title{EarSpy: Spying Caller Speech and Identity through Tiny Vibrations of Smartphone Ear Speakers}

\author{Ahmed Tanvir Mahdad}
\affiliation{%
  \institution{Texas A\&M University}
}
\email{mahdad@tamu.edu}

\author{Cong Shi}
\affiliation{%
  \institution{New Jersey Institute of Technology}}
\email{cs638@njit.edu}

\author{Zhengkun Ye}
\affiliation{%
  \institution{Temple University}
}
\email{zhengkun.ye@temple.edu}

\author{Tianming Zhao}
\affiliation{%
  \institution{University of Dayton}
}
\email{tzhao1@udayton.edu}

\author{Yan Wang}
\affiliation{%
  \institution{Temple University}
}
\email{y.wang@temple.edu}

\author{Yingying Chen }
\affiliation{%
  \institution{Rutgers University}
}
\email{yingche@scarletmail.rutgers.edu}

\author{Nitesh Saxena}
\affiliation{%
  \institution{Texas A\&M Unviersity}
}
\email{nsaxena@tamu.edu}

\begin{abstract}

Eavesdropping from the user's smartphone is a well-known threat to the user's safety and privacy. Existing studies show that loudspeaker reverberation can inject speech into motion sensor readings, leading to speech eavesdropping.
% Researchers have already explored the \textcolor{blue}{eavesdropping phone conversations} by leveraging loudspeaker reverberation and its effect on the smartphone's built-in \textcolor{blue}{motion sensors which do not require user's permission to access} during a phone conversation. 
% \textcolor{blue}{Recent research studies also reveal the potential of sensing such vibrations using dedicated radio frequency devices.} 
While more devastating attacks on ear speakers, which produce much smaller scale vibrations, were believed impossible to eavesdrop with zero-permission motion sensors.
% \textcolor{blue}{Prior works demonstrated the motion sensor's effectiveness in sensing speech features from a phone conversation, while we are using ear speakers, which produce much smaller scale vibrations.} 
In this work, we revisit this important line of reach. We explore recent trends in smartphone manufacturers that include extra/powerful speakers in place of small ear speakers, and demonstrate the feasibility of using motion sensors to capture such tiny speech vibrations. We investigate the impacts of these new ear speakers on built-in motion sensors and examine the potential to elicit private speech information from the minute vibrations. Our designed system \spy{} can successfully detect word regions, time, and frequency domain features and generate a spectrogram for each word region. We train and test the extracted data using classical machine learning algorithms and convolutional neural networks. We found up to 98.66\% accuracy in gender detection, 92.6\% detection in speaker detection, and 56.42\% detection in digit detection (which is 5X more significant than the random selection (10\%)). Our result unveils the potential threat of eavesdropping on phone conversations from ear speakers using motion sensors.

\end{abstract}

\keywords{Eavesdropping, ear, Motion Sensor} % TODO: replace with your keywords

\maketitle

\pagestyle{plain}

\settopmatter{printfolios=true}

\input{body}

%\newpage
\newcommand{\BIBdecl}{\setlength{\itemsep}{0.25 em}}
\raggedright
\bibliographystyle{ACM-reference-format}
\bibliography{references}

\appendix

\end{document}

%% file: body.tex
\section{Introduction}
\label{sec:intro}
\input{introduction}

\section{Background}
\label{sec:background}

\input{background}

\section{Related Work}
\label{sec:related Work}
\input{related_work}

\section{Design and Implementation}
\label{sec:design}

\input{design}

\section{Evaluation}
\label{sec:evaluation}

\input{evaluation}

\section{Discussion and Future Work}
\label{sec:discussion}

\input{discussion}

\section{Conclusion}
\label{sec:conclusion}
\input{conclusion}

%% file: introduction.tex
\begin{figure}[t]
 \centering
        %\vspace{-5mm}
 \includegraphics[scale=.25]{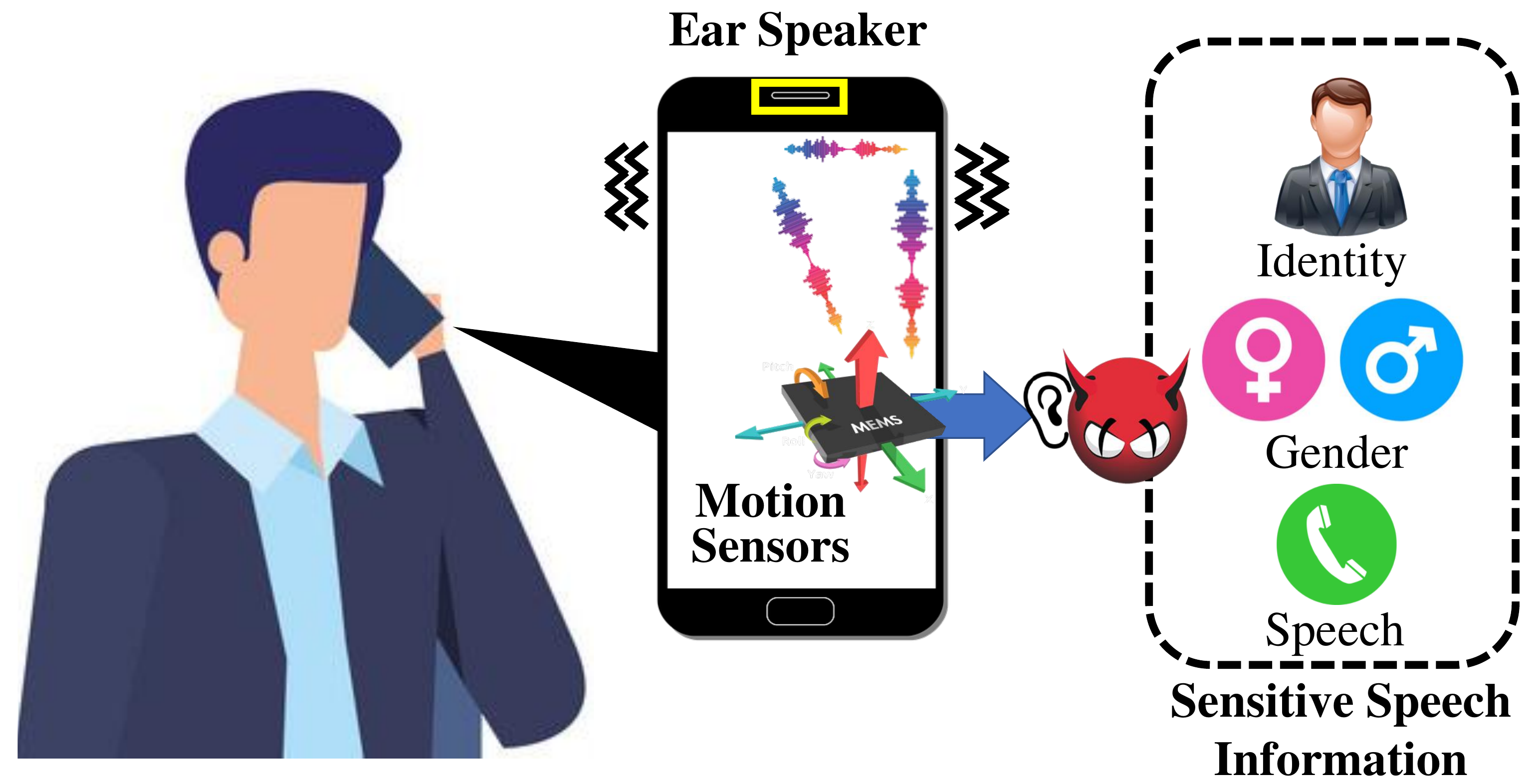}
        %\vspace{-5mm}
 \caption{Overview of ear speaker eavesdropping.}
 \label{fig:overview}
        %\vspace{-3mm}
\end{figure}

Eavesdropping on smartphones is always a well-known threat and a major security concern for users. Call recording is the most straightforward approach for an adversary to eavesdrop. However, smartphone operating systems are imposing restrictions on third-party apps for recording calls using microphones \cite{androidauthorityrecording,googlecallrecording}, which thwarts most attacks relying on microphone access.

A possible workaround for adversaries can be extracting speech information from zero-permission motion sensors through a side-channel attack. 
% Eavesdropping through smartphone sensors is a well-known threat, and 
It is a significant privacy concern that users are unaware of \cite{crager2017information} but have been extensively investigated by researchers in the last decade. Researchers have reported potential eavesdropping prospects using motion sensors \cite{anand2019spearphone,ba2020learning,hu2022accear}, keystrokes on touchscreens \cite{yu2019indirect}, stylus pen writing \cite{liu2020maghacker} and using external devices \cite{wang2022mmeve,basak2022mmspy}. Furthermore, eavesdropping through light sensors \cite{chakraborty2017lightspy}, gyroscope \cite{michalevsky2014gyrophone} are also reported in the literature.

Among the built-in sensors of smartphones, motion sensors are mostly known as vulnerable to eavesdropping. Adversaries leverage motion sensors to collect audio (e.g., voice conversation \cite{ba2020learning}), touch screen inputs \cite{xu2012taplogger}, and even indoor locations \cite{zheng2019missile}. Eavesdropping through motion sensors is straightforward, as adversaries do not need explicit permission to collect raw data from them.

\begin{figure*}[!ht]
%\vspace{-7mm}
 \subfloat[Spectrogram generated from accelerometer data of Oneplus 3T ear speaker (older model, no stereo speakers). \label{fig:spectrogram-3t}]{%
 \includegraphics[width=0.32\textwidth, height=4cm]{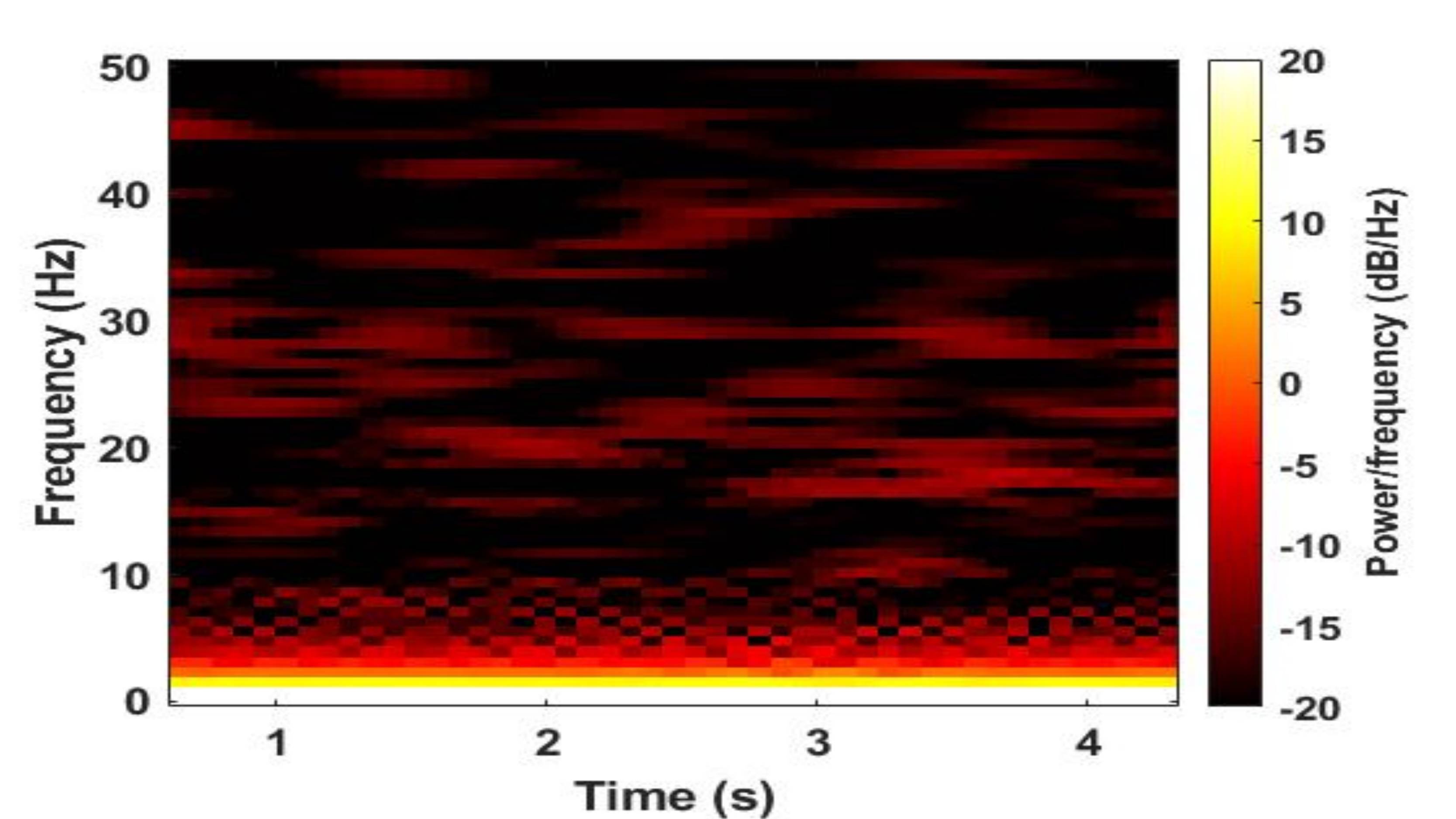}
 }
 \hfill
 \subfloat[Spectrogram generated from accelerometer data of Oneplus 7T ear speaker (newer model, with stereo speakers).\label{fig:spectrogram-7t-ear}]{%
 \includegraphics[width=0.32\textwidth, height=4cm]{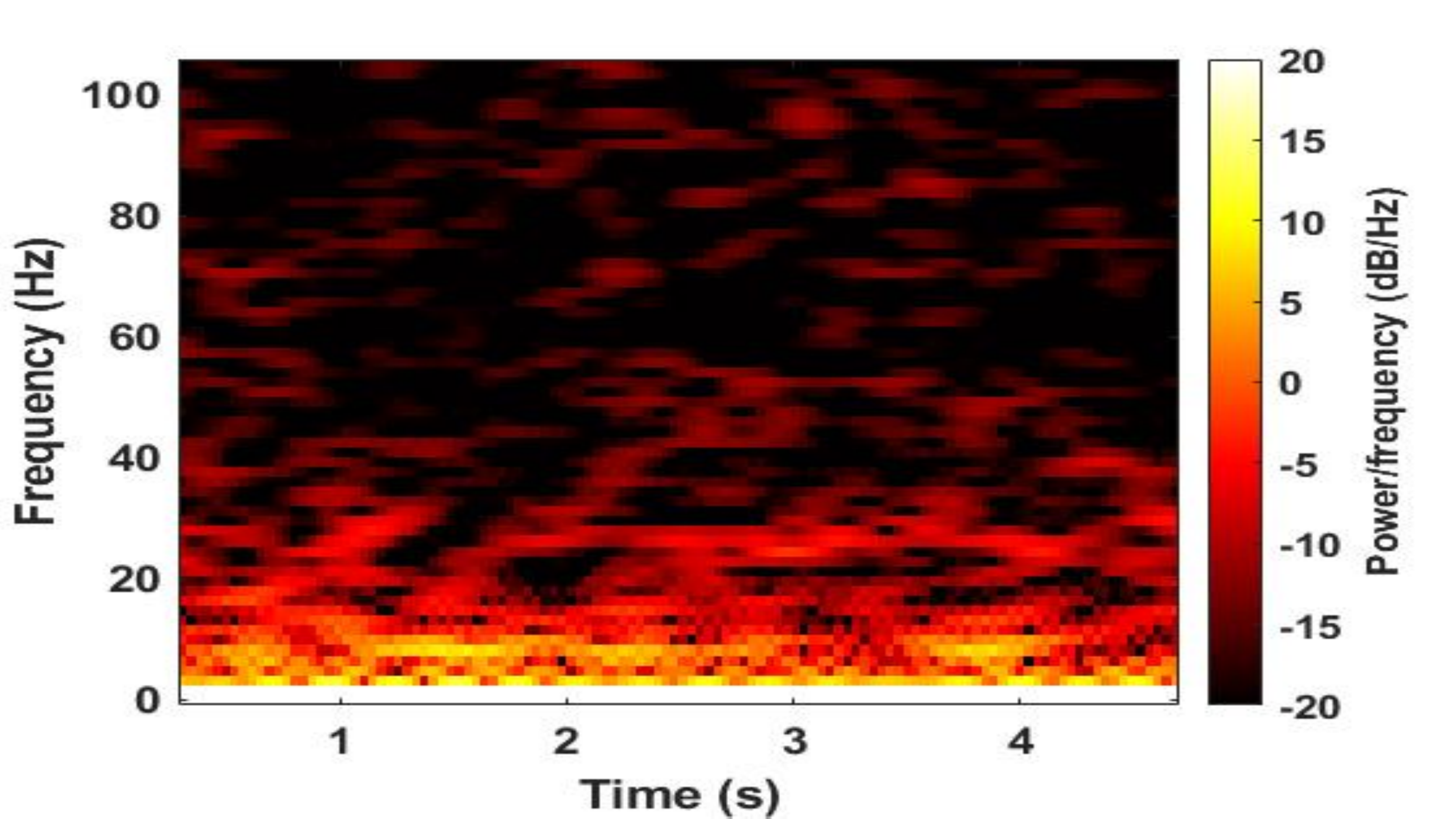}
 }
 \hfill
 \subfloat[Spectrogram generated from accelerometer data of Oneplus 7T loud speaker (newer model, with stereo speakers).\label{fig:spectrogram-7t-loud}]{%
 \includegraphics[width=0.32\textwidth, height=4cm]{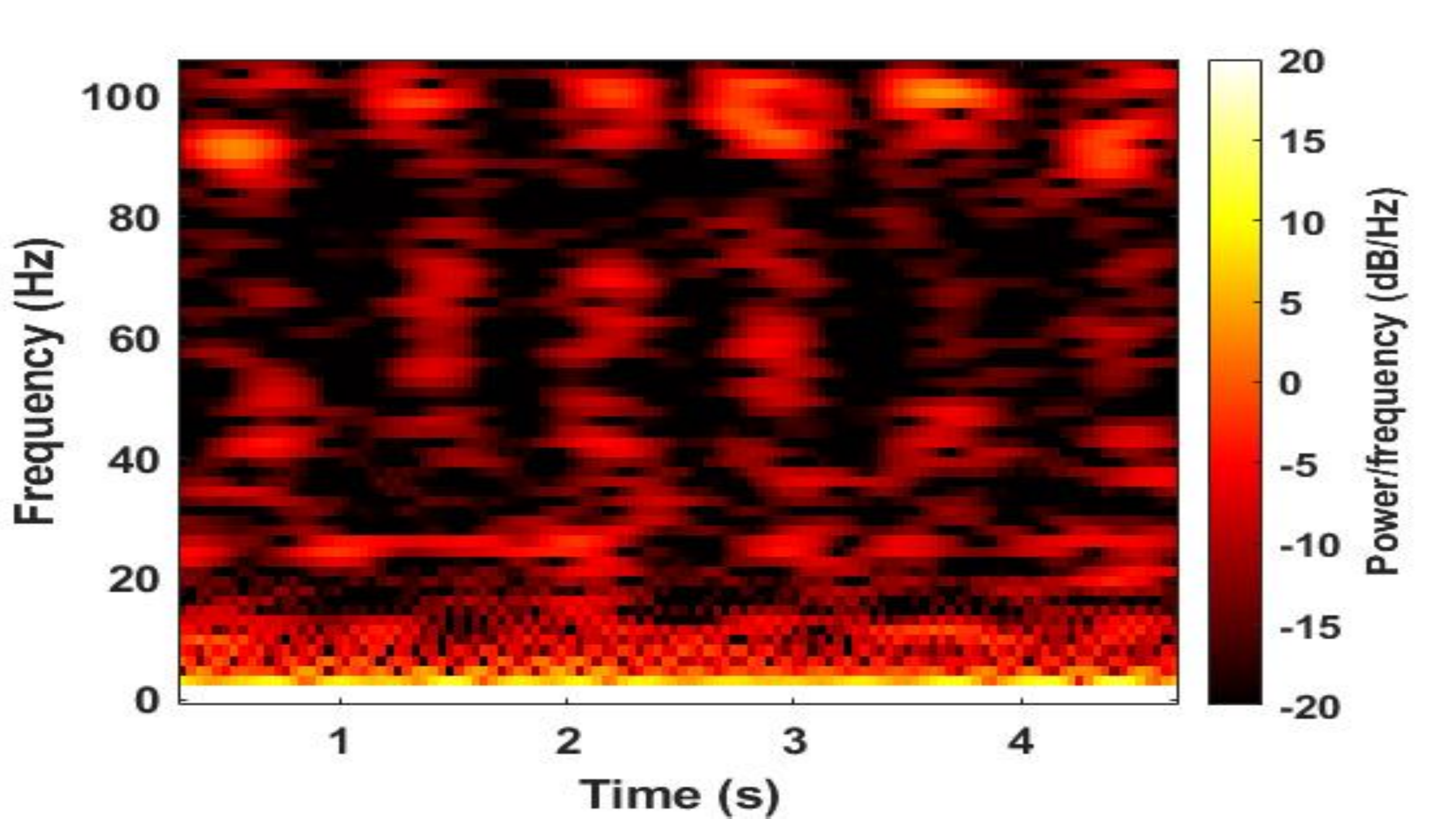}
 }
 
 	%\vspace{-2mm}
 \caption{Spectrogram generated while playing word ``Zero" for six times.}
\label{fig:spectrogram-comparison}
  \vspace{-3mm}
 	
 \end{figure*}

An abundant amount of work has been done on eavesdropping attacks induced by vibration generated from phone loudspeakers (e.g., \cite{anand2019spearphone,ba2020learning}). However, very few works have been done on eavesdropping \textit{ear speakers}, a built-in internal speaker in a smartphone that is used to listen to the conversation while the phone is held to the ear. Eavesdropping on the ear speaker is the most practical attack vector that can eavesdrop on phone conversations, as most people are not willing to expose sensitive speech, especially in public places. A few recent studies~\cite{wang2022mmeve, basak2022mmspy} show that the vibrations produced by ear speakers can be captured using high-resolution wireless sensors placed close to the victim.
% Wang et al. used commercial external mmWave sensors to determine the ear speaker voice in their work mmEve \cite{wang2022mmeve}. In a similar work, \cite{basak2022mmspy}, Basak et al. used off-the-shelf car-mounted radar devices to eavesdrop on a phone conversation from the ear speaker.

A natural question is that: Is it possible to eavesdrop on ear speakers using built-in motion sensors? Such an attack setting is highly practical due to the zero-permission property of motion sensors, which does not require placing or hacking any devices in the victim's environment. Previous studies did not find enough impact of ear speakers on the accelerometer (e.g., Figure 10 of \cite{anand2019spearphone}). However, we find that the audio quality of smartphone speakers continues to improve and evolve \cite{speakerEvolution}. Following the trend, recent flagship smartphones contain stereo speakers, which requires placing two speakers at the top and bottom. In most cases, traditional small ear speakers are replaced by more prominent stereo speakers. As a result, phones with stereo speakers produce more sound pressure during conversations than phones with conventional ear speakers. In Figure \ref{fig:spectrogram-comparison}, a comparative spectrogram analysis of two smartphones (Oneplus 7T contains stereo speakers, whereas OnePlus 3T does not) presents a noticeable difference in its vibration effect in a motion sensor (i.e., accelerometer) while playing a recording of the word ``Zero" six times in five seconds interval. Figure \ref{fig:spectrogram-3t} shows a spectrogram demonstrating the very low impact of ear speakers in the accelerometer in an older model phone (OnePlus 3T) where stereo speakers were not present. Figure \ref{fig:spectrogram-7t-ear} shows some impact on the accelerometer due to vibration induced by ear speakers of a newer model smartphone (i.e., OnePlus 7T) compared to Figure  \ref{fig:spectrogram-3t}. Figure \ref{fig:spectrogram-7t-loud} shows the spectrogram for loudspeakers of the OnePlus 7T with a clear view of word regions.

Based on these observations, we propose to analyze the accelerometer data and try to extract sensitive speech and speaker information from a speech played on ear audio. Although phone manufacturers use a larger and more powerful speaker at the top in place of ear speakers, during a phone conversation, the volume is controlled at a level so that users do not experience any discomfort. We use public speech datasets (e.g., Free Speech Data Set \cite{FSDD}, JL-Corpus \cite{jlcorpus}, emo-DB \cite{emodb}) in our experiment, and our word region detection program can still detect more than 50\% of ``word region" from the raw accelerometer data. We extract time and frequency domain features, generate spectrograms, and use classical machine learning algorithms and deep learning techniques to examine if they can detect speech information (e.g., words, speakers, gender) from accelerometer data. Our analysis reveals that an adversary can successfully reveal Gender, Speaker, and speech information with reasonable accuracy (98\% for Gender detection, 92\% for speaker detection). An overview of the system is illustrated in Figure \ref{fig:overview}.

\noindent\smallskip\textbf{Our Contribution}: We analyze the effect of vibration induced by ear speakers during a conversation, extract time-frequency domain features, and generate a spectrogram for each identified word region. We use classical machine learning and deep learning techniques to identify the speech, speaker, and gender of the caller by analyzing the features. Our contribution to this work is three-fold:

\begin{enumerate}

\item\textit{\textbf{Exploration of Eavesdropping Opportunity on Ear Speaker using Built-in Motion Sensors}}:  Eavesdropping from ear speakers is one of the most real-world and practical threats. Researchers have already explored this area \cite{wang2022mmeve,basak2022mmspy} in the previous literature that uses external radars/devices outside of the smartphone device. However, to the best of our knowledge, \spy{} is the first work that explores the eavesdropping opportunity on ear speakers using built-in motion sensors of recent smartphones with stereo speakers.

\item\textit{\textbf{Extraction of Word Regions and Features from Accelerometer Data}}: Although we observe a little impact of ear speaker-induced vibration in accelerometer data,  we are able to identify more than 45\% word regions to analyze. We also extract time and frequency domain features and generate a spectrogram for each identified word region. We use these features and spectrograms to feed into classical machine learning and deep learning techniques to further analyze the accuracy of speech, speaker, and gender detection.

\item\textit{\textbf{Achieved Reasonable Accuracy in Detecting Speech and Speaker Specific Information}}: We achieve high accuracy in detecting speech (56\% accuracy) and speaker information (i.e., speaker (92\% accuracy) and gender identification (98\% accuracy)). Compared to audio and vibration domain (using loudspeaker) performance, this result is promising and reveals the real-world threat of voice conversation eavesdropping. 

\end{enumerate}

%% file: background.tex
%\redtext{\smallskip\noindent\textbf{Impact of Powerful Earphone Speakers of Smartphone (Not sure this original disscussion paragraph should be added here)}: Earpiece speakers are designed to produce low-volume sound during phone conversation. The volume level is set to an optimum level so that it can be comfortable to hear during a conversation in the handheld position. Smartphone speakers are evolving fast in the last decade \cite{speakerEvolution}, and the recent trend is to introduce stereo speakers with smartphones. Generally, stereo speakers are two speakers built into the top and bottom portions of phones. As such, manufacturers are designing phones with better quality speakers at the top, which is also used as an earpiece speaker during the phone conversation.}

\begin{figure}[t]
 \centering
        %\vspace{-5mm}
 \includegraphics[height=4cm, width=.48\textwidth]{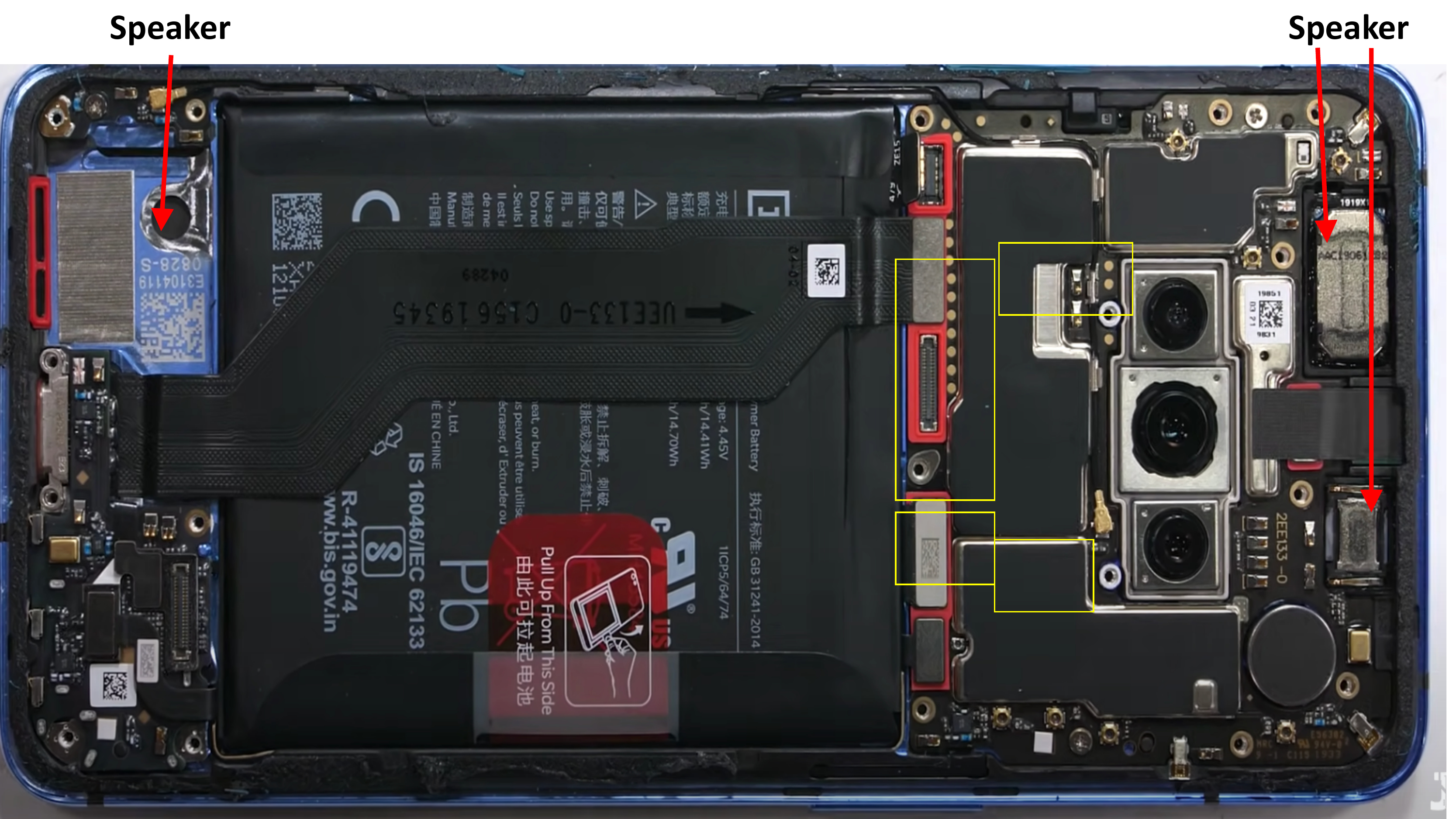}
        %\vspace{-5mm}
 \caption{Teardown snapshot \cite{op7tteradown} of OnePlus 7T.}
 \label{fig:op7T}
        %\vspace{-3mm}
\end{figure}

%Although some phone manufacturers claim that their phone has stereo speakers, their designed top speakers are not as powerful as the bottom primary loudspeaker. 
\smallskip\noindent\textbf{Ear Speakers on Smartphone.} Ear speakers are designed to produce low-volume sound during phone conversation where the user places the phone against his ear in order to clearly hear the sound from the ear speakers. Fig~\ref{fig:op7T} depicts the layouts of ear speakers on a typical phone model (i.e., Oneplus 7T). Specifically, the speaker at the bottom is typically the loudspeaker. The ear speakers of a smartphone is mounted on the top area of the smartphone’s motherboard. Since the vibrations generated by the ear speaker are much weaker than the loudspeaker, therefore, a direct contact between the ear speaker and the user's ear is ideal for high-quality sound reception. The main reason why it helps is that sounds propagating among two solid surfaces (i.e., ear speaker and ears) are much better than no physical contact case (i.e., air as the intermediate medium).

\smallskip\noindent\textbf{Vibration Captured by Motion Sensor on Smartphone.} Modern smartphones are equipped with highly sensitive  motion sensors (i.e., accelerometer and gyroscope) that are designed for sensing phone vibrations. Existing studies~\cite{anand2021spearphone,su2021towards} have shown that the vibration of the phone body caused by the transmitted sound from the built-in speaker can be captured by the motion sensor. The basic principle is that the sound transmitted through the smartphone's body generates vibrations, and the motion sensor on that smartphone can capture those vibrations. More specifically, Spearphone~\cite{anand2021spearphone} found that the accelerometer on the smartphone has a strong response to the sound frequency from 100Hz to 3300Hz. Moreover, they observed that sounds at different frequencies generate responses at the low-frequency points of the accelerometer, known as aliased signals. And it can be expressed using the equation as follows: $f_a=\left|f-N \cdot f_s\right|$, where $f_a$ is the vibration frequency of the accelerometer, $f$ is the sound frequency, $f_s$ is the accelerometer sampling rate, and $N$ can be any integer. This effect shows that the accelerometer can capture rich information in low-frequency aliasing signals since they are derived from the original sound at different frequencies. In addition, they also compared the frequency response of both accelerometers and gyroscopes and found that the accelerometer’s response was stronger than the gyroscope’s response in the frequency range 100Hz to 3300Hz. Therefore, we only adopt accelerometers in our experiments as well. 

%% file: related_work.tex
% We separate the related work into three parts based on the eavesdropping target: (1) direct sensing on the sound source using wireless signals (require deploying devices close to the loud- speaker/human subjects); (2) indirect sensing on sound-induced vibrations on objects motion sensors (lasers, wireless signals, motion sensors), wireless signals (need loud sound volume); (3) speech eavesdropping through built-in motion sensors (which is closest to our work).

\smallskip\noindent\textbf{Direct speech sensing on external loudspeakers/human throats.} Many research investigations have extended the search for speech eavesdropping from using a tampered/hidden microphone to radio frequency (RF) sensors, such as WiFi~\cite{wang2014we, wei2015acoustic}, Ultra-Wideband~\cite{wang2020uwhear}, and mmWave signals~\cite{xu2019waveear}. For example, researchers have explored using WiFi signals to recover the sound of speaker devices~\cite{wei2015acoustic} and capture mouth motions~\cite{wang2014we} for speech eavesdropping. These attacks rely on external and potentially customized/dedicated sensing devices around the human subjects for sensing, rendering these attacks cumbersome and less stealthy.

\smallskip\noindent\textbf{Indirect speech sensing based on vibrations.} Speech eavesdropping is also shown feasible through sensing sound-induced vibrations using various types of sensors~\cite{michalevsky2014gyrophone, zhang2015accelword, walker2022laser, muscatell1984laser, davis2014visual}. Gyrophone~\cite{michalevsky2014gyrophone} first showcased the attack setup where a smartphone is placed on the same solid surface as a loudspeaker. The smartphone’s gyroscope is then used to capture the surface vibrations induced by the speech playbacks of the loudspeaker. Recent studies further demonstrate the feasibility of such attacks through vibration sensing using lasers, high-speed cameras, and light sensors. For example, Davis et al.~\cite{davis2014visual} utilizes a high-speed camera to capture video streams to recover vibrations from some room objects (e.g., a bag of chips). Nassi et al.~\cite{nassi2021glowworm} show that sound vibrations on lamps can be detected and recovered by using electro-optical sensors. These attacks are promising, but they require a loud sound volume of the loudspeaker (e.g., 70$\sim$110dB) or a close distance between the vibration surface and the loudspeaker to trigger surface vibrations. Differently, this work targets more realistic attack scenarios, where the sounds with a low sound volume of around 50dB are produced by the ear speaker of smartphones. Compared to these existing approaches, our attack is also more resilient to impacts of environments, such as the occlusion by walls and movements of nearby human subjects.

\smallskip\noindent\textbf{Speech eavesdropping on smartphone speaker.} Instead of using external sensors, Spearphone~\cite{anand2019spearphone} and AccelEve~\cite{ba2020learning} recently demonstrated new eavesdropping attacks which derive speech based on the motion sensors on the same smartphone. The vibrations produced by the speaker can propagate through the motherboard and reach the motion sensors. With the motherboard as the vibration medium, it is more stable for the motion sensors to pick up speech vibrations. AccEar~\cite{hu2022accear} takes one step forward to design a deep neural network to reconstruct audio signals from the motion sensor readings. The attacks show promising results, but they assume the sounds are relayed by built-in loudspeaker, which is audible to nearby people and is less likely to use for phone calls in public spaces (e.g., offices, conferences). Different from these prior works, our attack targets minute speech playback by an ear speakers, which is completely inaudible to nearby human subjects. Our attack is more devastating as users normally believe the confidentiality of the speeches played via ear speakers (e.g., one-time passwords, birthdays, and social security card numbers) has been enforced.

% \bluetext{Tanvir: I think we can focus more on (3) as they are more related to our work. Also, if you suggest that we should include (1) and (2), you can also briefly discuss them. Also, should I add the above reference to bib file, so that, you can just discuss and cite right away?}

%% file: design.tex
We design a system that uses motion sensor data from the user's smartphone induced by ear speaker vibrations. Our goal is to examine if the ear speakers cause distinguishable vibration patterns on motion sensor data. This section discusses the system design and tools used for the experiment in detail. 

\subsection{Feasibility Determination}

\smallskip\noindent\textbf{Playing Voice Through Ear Speakers}:  We have used some third-party Android apps (Ear \cite{earpiece}, Mobile Ear Speaker Earphone \cite{mobileEarphone}) to play audio only through the ear speakers. We have used available, and widely known voice data sets (e.g., JL-Corpus \cite{jlcorpus}), FSDD \cite{FSDD}, Emo-DB \cite{emodb}) and played the audio through ear speakers.

\smallskip\noindent\textbf{Impact of Powerful Ear Speakers of Smartphone}: Ear speakers are designed to produce low-volume sound during a phone conversation. The volume level is set to an optimum level so that it can be comfortable to hear during a conversation in the handheld position. Smartphone speakers are evolving fast in the last decade \cite{speakerEvolution}, and the recent trend is to introduce stereo speakers with smartphones. Generally, stereo speakers are two speakers built into the top and bottom portions of phones. As such, manufacturers are designing phones with better quality speakers at the top, which is also used as ear speaker during phone conversations.

Although some phone manufacturers claim that their phone has stereo speakers, their designed top speakers are not as powerful as the bottom primary loudspeaker. We have analyzed some publicly available teardown videos \cite{op7tteradown, pixelteardown} and noticed some phone manufacturers use larger multiple speakers at the top to boost the audio quality (e.g., OnePlus 7T teardown at Figure \ref{fig:op7T}) whereas others use smaller speakers (e.g., Google Pixel 5 \cite{pixelteardown}). Smartphone with larger and multiple speakers is more likely to generate more vibration than smaller ones.

To test this hypothesis, we play a recorded word ``Zero" six times in five second interval through ear speakers of a smartphone where large dual speakers are used (e.g., OnePlus 7T) and collect accelerometer readings. We extract the word region and generate a spectrogram for it. We did the same experiment with another phone with less powerful speakers (OnePlus 3T). From the generated spectrogram (Figure \ref{fig:spectrogram-3t}, and Figure \ref{fig:spectrogram-7t-ear}), it is evident that audio played with a larger and improved speaker will cause distinguishable vibration, unlike previous phones.

\smallskip\noindent\textbf{Choosing Accelerometer to Collect Vibration Data}: We have already learned from previous literature \cite{anand2019spearphone} that the accelerometer performs better than a gyroscope to capture the vibration from the smartphone's internal speakers. So, in our experiments, we primarily focused on capturing accelerometer data.

\begin{figure}[!ht]
%\vspace{-7mm}
 \subfloat[Example of identified word regions from acceleromter data after applying 8Hz high-pass filter on loudspeaker. \label{fig:signal-loud}]{%
 \includegraphics[width=0.48\textwidth, height=3cm]{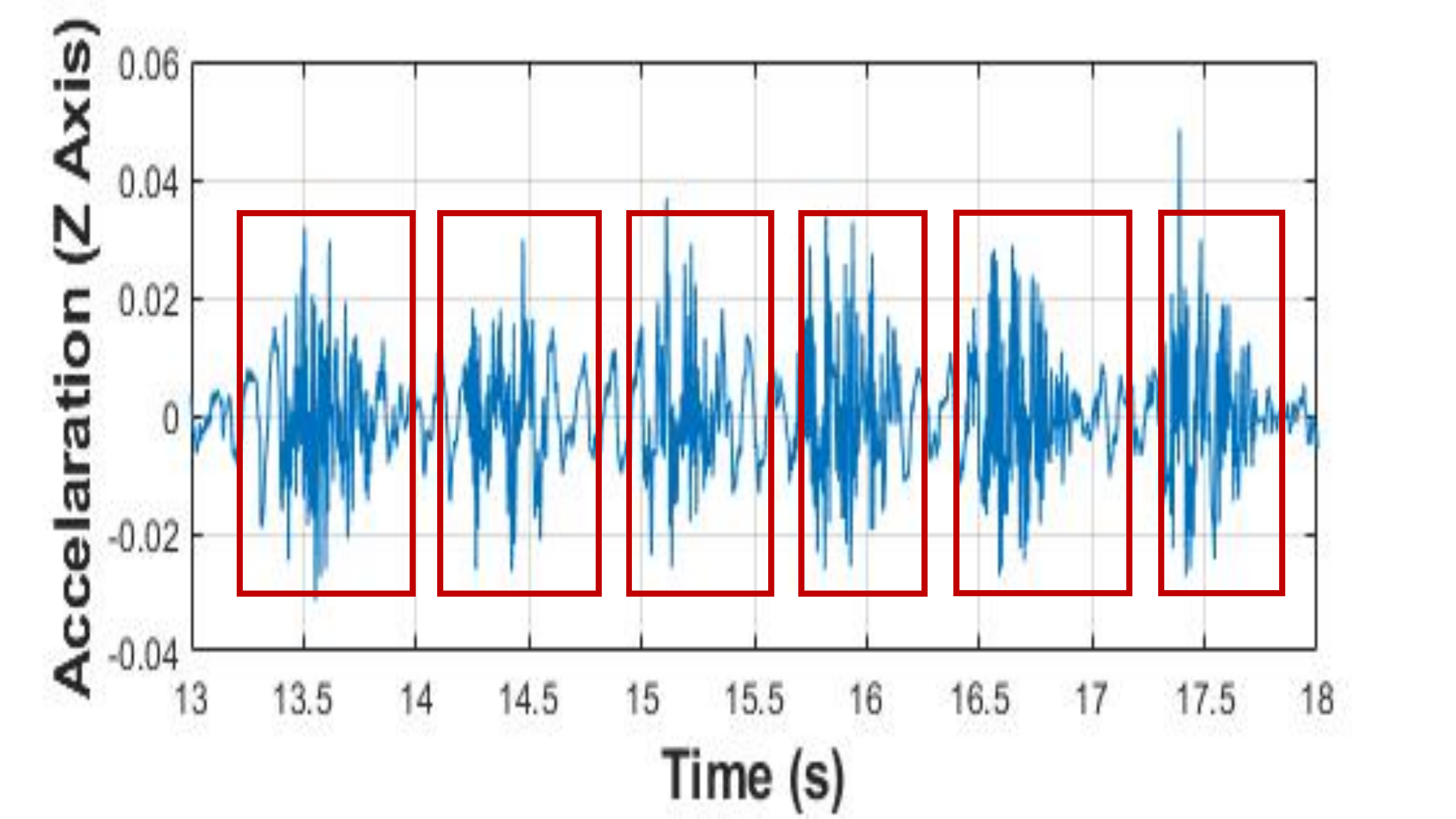}
 }
 \vfill
 \subfloat[Example of identified word regions from accelerometer data after applying 8Hz high-pass filter on ear speaker. \label{fig:signal-ear}]{%
 \includegraphics[width=0.48\textwidth, height=3cm]{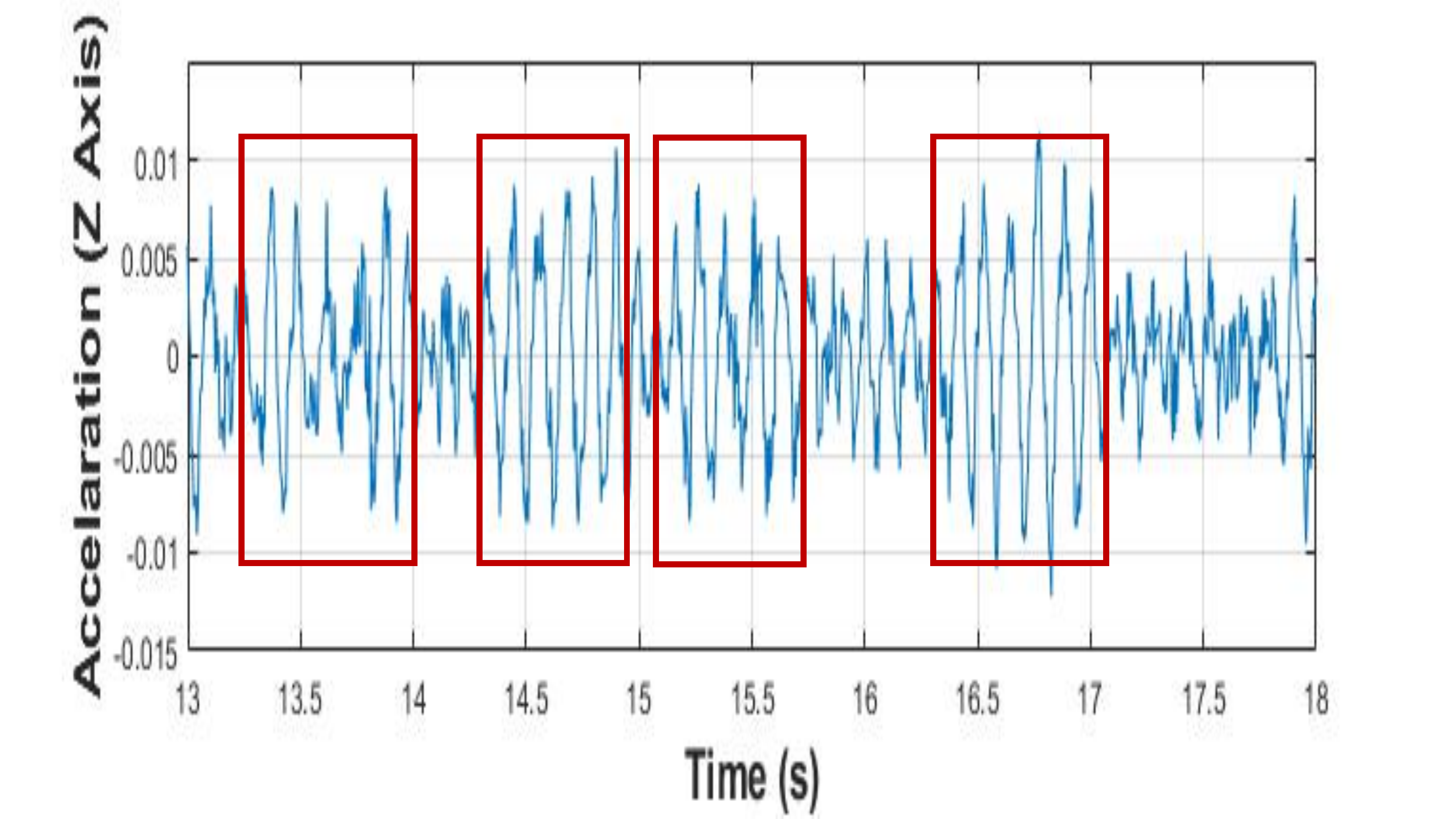}
 }
 	%\vspace{-2mm}
 \caption{Comparison of identified word regions from the loudspeaker and ear speaker setting.}
\label{fig:signal-comparison}
  \vspace{-3mm}
 	
 \end{figure}

\subsection{Word Region Identification}

As we discussed earlier, we play speech audio using publicly available speech datasets and collect the vibration data from the accelerometer. We place the phone in the handheld position (i.e., the natural posture of a human during a phone conversation). As a result, body and hand movements add low-frequency noises to the accelerometer data. We place a high-pass filter during our analyses to eliminate the effect of low-frequency body and hand movement.

% As we discussed earlier, 
Ear audio creates a very small impact on the accelerometer. So, if we set a larger value as a high-pass filter cutoff frequency, important speech features will be lost. Zhang et el., in their work accelWord \cite{zhang2015accelword}, also observed this challenge and did an information gain analysis to determine the optimum value. According to their analysis, if the cutoff value is greater than 2 Hz, then information gained with frequency domain features will reduce significantly. We did the information gain analysis on ear speaker data and found that even a value equal to or less than 1 Hz causes a significant amount of missing information. We collected all data from a single dataset in one go to avoid noise bias. 

After that, we analyze the vibration of speech generated in raw accelerometer data. In previous literature, \cite{anand2019spearphone,su2021towards}, authors claimed that they found the most impact of vibrations are generated by phone speakers along the Z-axis. As we are working with ear speakers, we measure the impact of tiny vibrations along the X, Y, and Z axis. We observe that variance along the X, Y, and Z axis as $1.7029*10^{-6}$, $1.7029*10^{-6}$, and $1.946*10^{-4}$. It is obvious that this observation is in line with the observation of previous literature, which implies the Z axis gets more impact of vibrations compared to the X and Y axis.

We developed a program in MATLAB to analyze the accelerometer data and detect the word region. When a speech is played on the ear speakers, spikes can be noticed in the Z-axis value of the accelerometer. We present accelerometer data when the word ``Zero" is uttered six times within 5 seconds timeframe in Figure \ref{fig:signal-comparison} for the loudspeaker and ear speaker scenario. As the loudspeaker has a larger impact on accelerometer data, all individual word regions are visible (Figure \ref{fig:signal-loud}) and easy to detect. In contrast, ear speaker has a lower impact on accelerometer data (\ref{fig:signal-ear}) and hence are hard to detect. In our presented example of Figure \ref{fig:signal-comparison}, where all six word regions are visible for the loudspeaker, only four word regions can be distinguished for the ear speaker. We observe that our program can automatically detect at least 45\% of word regions from the raw accelerometer data and calculate time and frequency domain features (detailed discussion in the following subsection).

\subsection{Tools Used for This Study}

As we have discussed before, our primary goal is to analyze the accelerometer data while playing audio from ear speakers. For this experiment, we choose smartphones that have powerful/ multiple ear speakers. We have used OnePlus 7T and the OnePlus 9, which meet the requirements. 

Both phones used for testing run on Android (Oneplus 7T runs on Android 11 while Oneplus 9 runs on Android 12). We have used a third-party Android app \textit{Mobile Ear Speaker Earphone} \cite{mobileEarphone} that runs a service that redirects all the output audio through ear speakers with default volume. We have also used another third-party app \textit{Physics Toolbox Sensor Suite} \cite{physicsToolBox} to collect accelerometer data while audio from datasets is played. 

We have used a MATLAB program to analyze the accelerometer data and extract time and frequency domain features. To train the time and frequency features of data, we used \textit{Weka} \cite{weka}, which provides a collection of machine learning algorithms and essential analysis tools. We have also designed a Convolutional Neural Network (CNN) to analyze time and frequency domain data. 

Using our developed MATLAB program, we also generate a spectrogram for each word region. We have also designed a CNN-based image classifier that can be fed with the generated spectrograms and classify them to detect gender, speaker, and speech.

\subsection{Time-Frequency Domain Feature Analysis}

\begin{table}[t]
\begin{center}
\caption{Time and frequency domain features.}
    \begin{tabular} { | p{4cm} | P{4cm} |  }
    \hline
     \textbf{Time Domain Features} & \textbf{Frequency Domain Features}  \\
    \hline
    minfreq, maxfreq, meanfreq, standard deviation, variance, range, CV, skewness, kurtosis, quantile25, quantile50, quantile85, maeanCrossingRate & Energy, Entropy, Frequency Ratio, Irregularity K, Irregularity J, sharpness, smoothness, specCentroid, specStdDev, specCrest, specSkewness, specKurt\\
    \hline
    \end{tabular}
\label{tab:freq}
\end{center}
\end{table}

We extracted time and frequency domain features using our developed MATLAB program. We use these features to train classical machine-learning algorithms using Weka. Initially, we checked with 40 different classifiers and found \textit{RandomForest, RandomSubspace, and DecisionTables} are showing better performance than others. We use 80\%/20\% train/test split and 10-Fold cross-validation. We also use these time-frequency domain features to train our developed CNN model, and there we have also used 80\%/20\% train/test splits. The time-frequency domain features we have used for this experiment are listed in Table \ref{tab:freq}.

\subsection{CNN Model Details}
\subsubsection{\textbf{Spectrogram-based Image Classifier}}

\begin{figure}[t]
 \centering
        %\vspace{-5mm}
 \includegraphics[scale=.25]{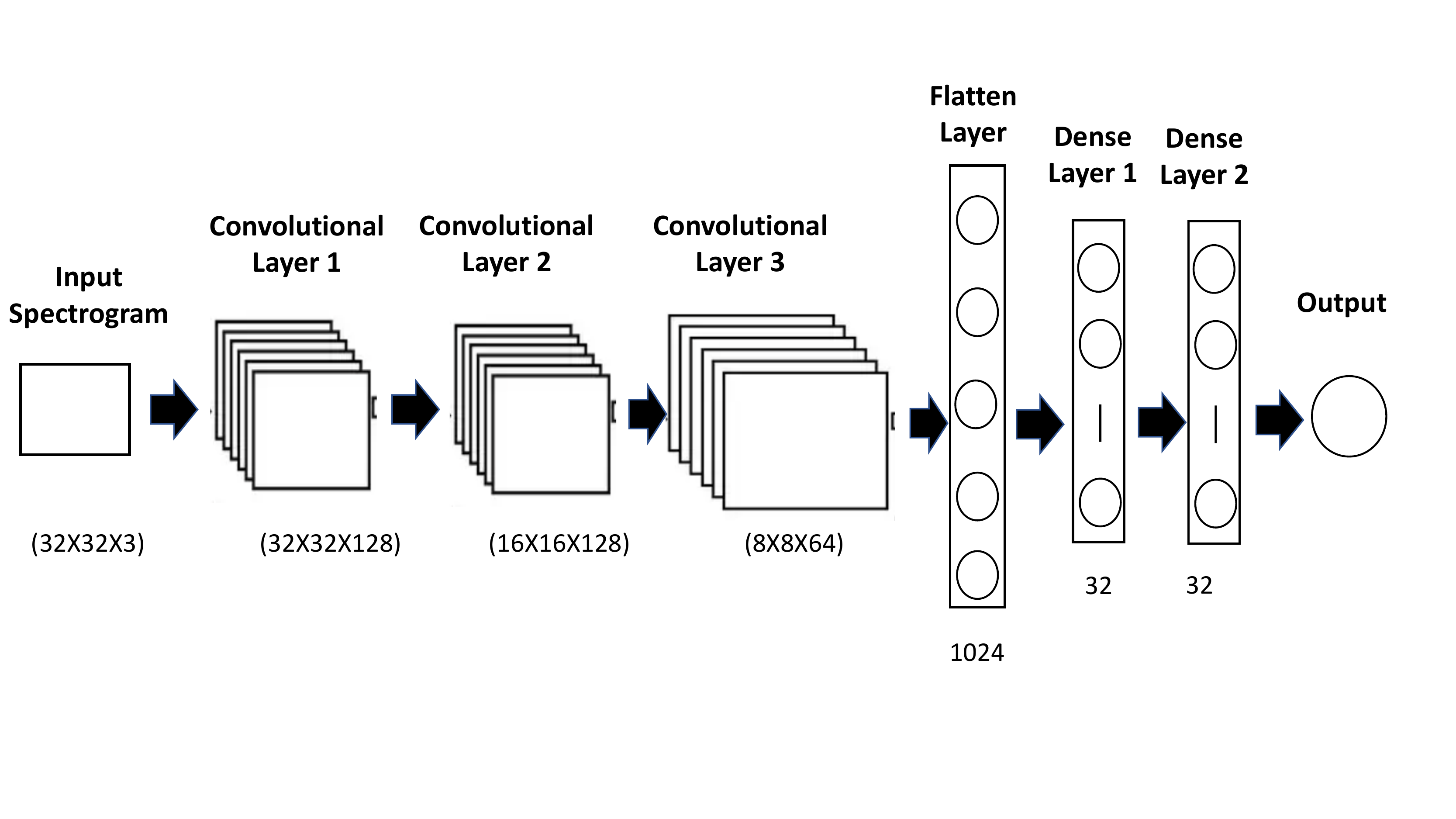}
        %\vspace{-5mm}
 \caption{CNN model used for spectrogram-based image classifier.}
 \label{fig:cnnspec}
        \vspace{-3mm}
\end{figure}

We use an image classifier for spectrogram analysis to take spectrogram inputs and then classify them.

\smallskip\noindent\textbf{Pre-processing}: We prepare training and testing data from the generated labeled spectrograms as a Hierarchical Data Format version 5 (HDF5) file. Afterward, we convert the generated spectrogram into 128X128 images and prepare training and testing data by attaching appropriate labels.

\smallskip\noindent\textbf{CNN Details}: In our designed model, there are three convolutional layers followed by three fully connected layers shown in Figure \ref{fig:cnnspec}. The first convolutional layer takes 128X128 images and contains 128 filters. The second and third convolutional layer includes 128 and 64 filters, respectively. Each convolutional layer is followed by a ReLU function, a dropout layer with 0.2 rates, and a max-pooling layer (pool size (2X2)). After three convolution layers, we placed three fully connected layers. The first two layers reduce the size of the image to 128 and 64, respectively, with the ReLU activation function. The third layer comes up with the ``softmax" activation function and changes the image size according to class size. The detailed CNN model is illustrated in Figure \ref{fig:cnnspec}. We have used Root Mean Square Propagation (RMSProp) optimizer while training the model with spectrogram images.

\subsubsection{\textbf{Time-frequency Domain Feature based CNN Classifier}}\label{cnn-tf}

We collect time-frequency domain features, write them into a CSV file and feed this data into our designed CNN model that classifies based on the time-frequency domain features. 

\smallskip\noindent\textbf{Pre-processing}: Our developed MATLAB program calculates time-frequency domain features with the label for each word region and generates a CSV file that contains all the information. After importing the CSV file, we check if there is any NaN (Not a Number) value on time and frequency domain features.

\begin{figure}[t]
 \centering
        %\vspace{-5mm}
 \includegraphics[scale=.25]{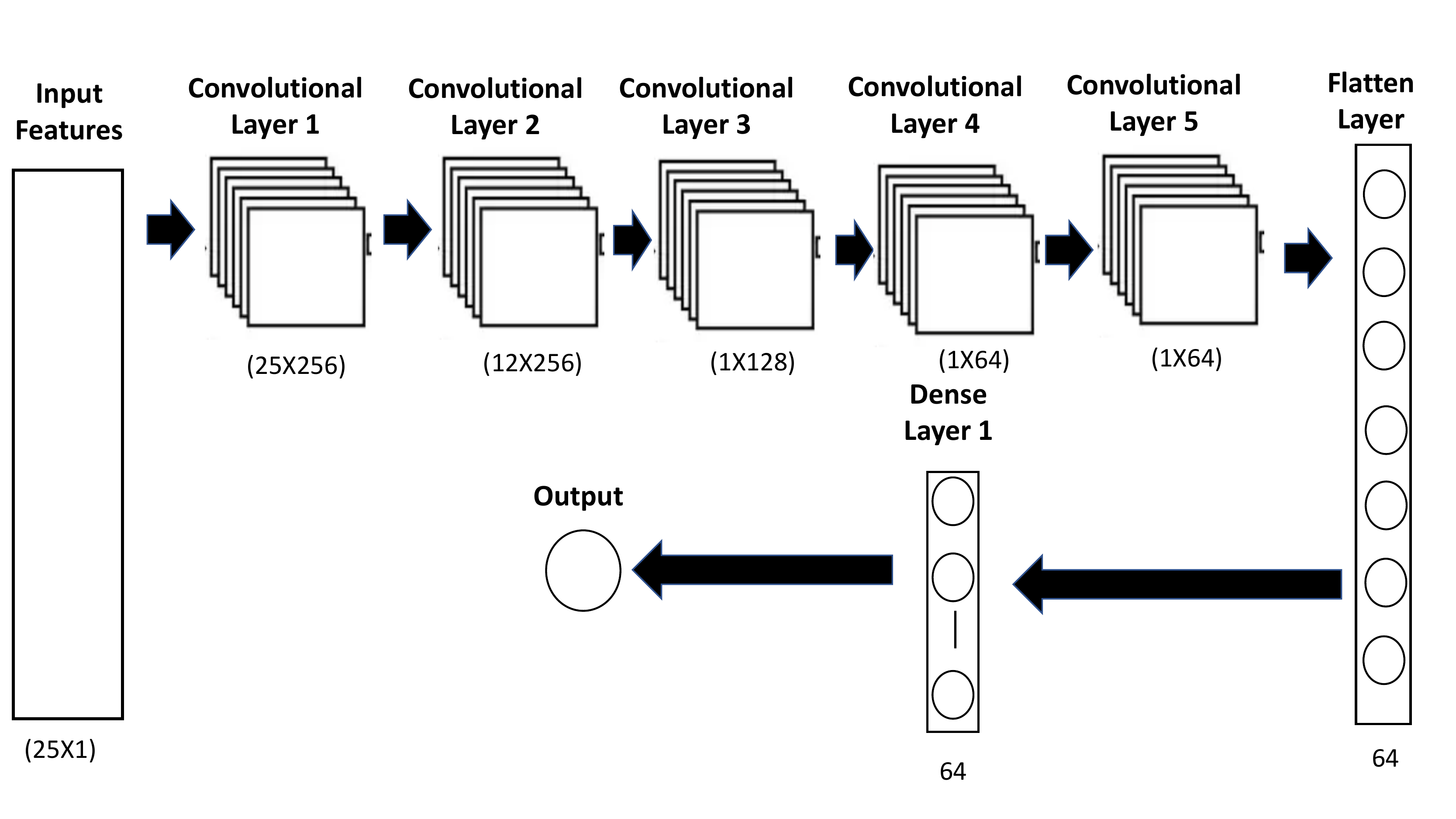}
        %\vspace{-5mm}
 \caption{CNN model used for time and frequency domain feature Analysis.}
 \label{fig:cnnfreq}
        \vspace{-3mm}
\end{figure}

\smallskip\noindent\textbf{CNN Details}: In our designed time-frequency feature-based classifier model, we use five convolutional layers followed by one dense layer (with softmax activation function). The first two convolutional layers contain 256 filters, the third convolutional layer has 128 filters, and the fourth and fifth one has 64 filters each. We have used dropout layers with a rate of 0.25 in the second and third convolution layers. We also used batch normalization in the second and third convolutional layers. Each of the convolutional layers used the ReLU activation function. Finally, we used a fully connected layer with a class size containing the softmax function. We have used Root Mean Square Propagation (RMSProp) optimizer while training the model with time and frequency domain features. An overview of the CNN model used for this purpose is depicted in Figure \ref{fig:cnnfreq}.

%% file: evaluation.tex
In this section, we discuss details about our experiment setup, dataset, and data collection methods. Most importantly, we discuss how our designed system can extract speech properties (i.e., speech, gender, and speaker information) from vibration induced by the ear speaker and evaluate the performance.
\subsection{Experiment Setup}

\smallskip\noindent\textbf{Data Collection Method}: We use the natural handheld position of smartphone users that they use during a phone conversation in the experiment. As discussed earlier, we play the audio from the selected dataset, and a third-party app collects the accelerometer data at that time.

\smallskip\noindent\textbf{Dataset Selection}: We use publicly available and well-known datasets to evaluate if our designed system can identify the speech information from the vibration induced by ear speakers. For gender and speaker detection, we use JL-Corpus \cite{jlcorpus}, and emo-DB \cite{emodb} datasets. The JL-Corpus dataset has 2400 utterances with four different speakers (two males and two females), whereas the emo-DB dataset has 535 utterances with ten actors (five males and five females). Actors use English in the JL-Corpus dataset and German in the emo-DB dataset as utterance language. For speech recognition, we use the digit dataset \textit{Free Spoken Digit Dataset} \cite{FSDD} with the utterance of six(6) actors. Each actor utters each digit ten times (a total of 500 utterances per actor). 

We have done a preliminary check on the datasets to examine the audio quality of the utterances. We have found that, in the FSDD dataset, three of the actor's data contain too much background noise or inconsistent volume during the recording. So, we removed these 1500 data and worked with only the remaining 1500 data in FSDD datasets. We have observed that data from emo-DB and JL-Corpus do not have similar problems. In addition to that, as the ear speakers produce low-volume audio output, the impact on the accelerometer is minimal. So, our word region detection program cannot identify all the word regions. However, it can detect 45\% to 90\% data, which is reasonable considering the low impact of vibration in the accelerometer. 

\smallskip\noindent\textbf{Device Selection}: As discussed earlier, our primary focus is to determine if ear speakers of recent smartphones that use stereo speaker feature and have powerful and multiple speakers on top are generating enough vibration on the motion sensor (i.e., accelerometer), so that, individual speech features can be identified. So, we measure the sound pressure level of each phone when they are playing the same audio from the FSDD dataset. From our experiment, we observe that only the \textit{OnePlus 7T} and the \textit{OnePlus 9} generate greater sound pressure than other phones (Oneplus 7T shows 42-46 dB, where OnePlus 9 shows 40-44 dB). So, we select these two phones for further experiments.

\smallskip\noindent\textbf{Posture Selection}: In this work, our primary goal is to evaluate the risk of voice conversation in the phone through ear speakers. So, experiment data collectors keep their phones in the natural handheld posture while collecting data. All data are collected when data collectors sit on the chair. We have collected accelerometer readings for the whole dataset on one go to avoid human movement noise-induced bias on different classes.

\subsection{Data Collection Details}

\smallskip\noindent\textbf{Audio File Preparation}: We collect all dataset's audio files and sort them according to class in a folder so that it plays one class after another. Before starting the data collection, we played the audio file once to recheck if every audio file was playing correctly and note down the time when the audio file of a specific class was finished.

\smallskip\noindent\textbf{Tools Used in Data Collection}: We have used \textit{Physics Toolbox Sensor Suite} to collect accelerometer data. The OnePlus 7T and the OnePlus 9 phones' default sampling rates are 420 Hz and 520 Hz, respectively. We collect accelerometer data using this sampling rate. We have exported the collected data as a CSV file which is used for the MATLAB feature extraction program's input.

\subsection{Feature Extraction}

We have developed programs using MATLAB for time and frequency domain feature extraction and spectrogram generation. The first program takes the accelerometer data CSV file and detects each word region, and then extracts the time and domain frequency feature of each word region. This program writes and exports all time and frequency domain features in an external file and labels the data according to class. After that,  we used Weka \cite{weka} to classify using classical machine learning algorithms.

The Spectrogram generator (developed using MATLAB) can also detect the word regions and generate spectrograms for each word region. The generated spectrograms are labeled according to class. Later these generated spectrogram is used to feed into our developed CNN for further analysis.

\subsection{Gender Recognition}

\begin{figure}[t]
%\vspace{-7mm}
 \subfloat[Gender recognition training loss Vs. validation loss for Emo-DB dataset (time-frequency feature analysis). \label{fig:gender-loss}]{%
 \includegraphics[width=0.22\textwidth, height=3cm]{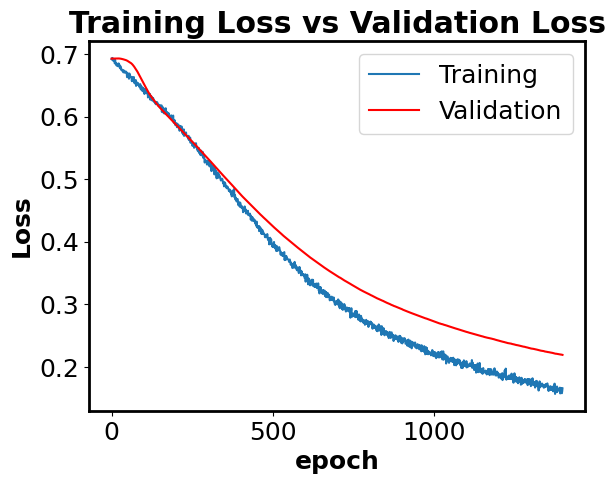}
 }
 \hfill
 \subfloat[Gender recognition training accuracy Vs. validation accuracy for Emo-DB dataset (time-frequency feature analysis). \label{fig:gender-acc}]{%
 \includegraphics[width=0.22\textwidth, height=3cm]{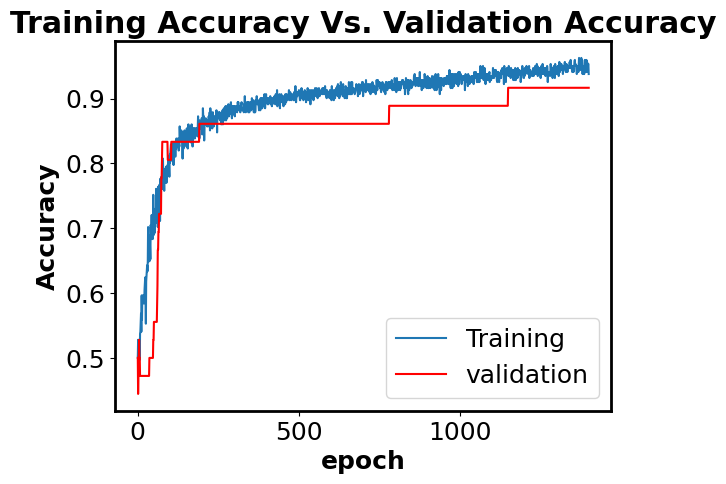}
 }
 \vfill

 \subfloat[Gender recognition training loss Vs. validation loss for Emo-DB dataset (spectrogram analysis). \label{fig:gender-loss-spec}]{%
 \includegraphics[width=0.22\textwidth, height=3cm]{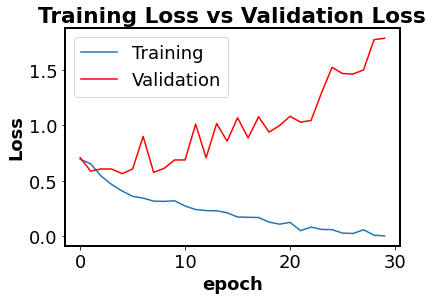}
 }
 \hfill
 \subfloat[Gender recognition training accuracy Vs. validation accuracy for Emo-DB dataset (spectrogram analysis). \label{fig:gender-acc-spec}]{%
 \includegraphics[width=0.22\textwidth, height=3cm]{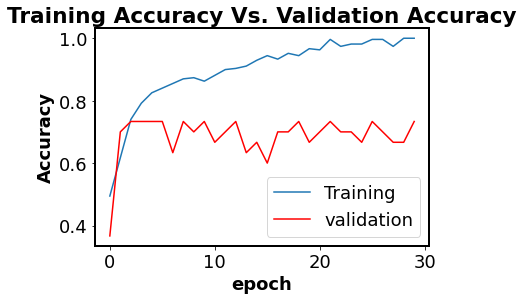}
 }
 	%\vspace{-2mm}
 \caption{Gender recognition training and validation accuracy graph using different methods.}
\label{fig:gender-chart}
  \vspace{-3mm}
 	
 \end{figure}

\smallskip\noindent\textbf{Gender Detection}: We use JL-Corpus \cite{jlcorpus}, and emo-DB \cite{emodb} dataset to evaluate if the caller's gender can be detected from the accelerometer data. We have used three methods to evaluate: (1) Classical machine learning algorithm with time and frequency domain features. (2) CNN with time and frequency domain features (3) CNN with generated spectrogram for each word region.

\smallskip\noindent\textbf{ML Algorithm with Time/Frequency Domain Features}: For the emo-DB dataset, our detection program can detect 448 word regions among 535 original utterances for the OnePlus 7T and 300 word regions for OnePlus 9. We have extracted all detected word regions' time/frequency domain features. We have used ``RandomForest", ``RandomSubspace", and ``Decision Table" as classifiers for our analysis. We have used 80/20 train/test split and 10-Fold cross-validation for our analysis.

For RandomForest Classifier, we have achieved 98.66\% accuracy in classifying genders. Whereas for RandomSubspace, we have also achieved 98.66\% accuracy, and for Decision Table, we have observed 98.21\% accuracy for the OnePlus 7T. The detailed result is shown in Table \ref{Gender} and Table \ref{result-op7}. Similarly, for the OnePlus 9, we get 88.67\%, 77.71\%, and 84.67\% accuracy for RandomForest, RandomSubspace, and Decision Table classifiers, respectively. The detailed result is shown in Table \ref{Gender} and Table \ref{result-op9}.

For the JL-Corpus dataset, our detection program can detect 1469 word regions among 2400 utterances. Here, we also have used RandomForest, RandomSubspace, and Decision Tree as classifiers for our analysis.

For RandomForest Classifier, we have achieved 78.62\% accuracy in classifying genders. Whereas for RandomSubspace, we have also achieved 79.37\% accuracy, and for Decision Table, we have observed 77.67\% accuracy for the OnePlus 7T. Similarly, we got 77.71\%, 74.20\%, and 72.14\% accuracy for the OnePlus 9. The detailed result is shown in Table \ref{Gender}, Table \ref{result-op7}, and Table \ref{result-op9}.

\smallskip\noindent\textbf{CNN with Time/Frequency Domain Features}: As discussed, we have extracted time and frequency domain features for 448 word regions for emo-DB and 1469 word regions for JL-Corpus datasets. We have designed a CNN to classify time and frequency domain features (Details are described in Section \ref{cnn-tf}).

We have used \textit{binary\_crossentroy} as loss function and \textit{Root Mean Square Propagation (RMSProp)} as the optimizer and 80/20 split as train/test split in our analysis. We have achieved 95.55\% for the emo-DB dataset and 75.71\% accuracy for the JL-Corpus dataset for the OnePlus 7T. We also got 83.33\% and 67.52\% accuracy for the emo-DB and JL-Corpus datasets, respectively, for the OnePlus 9 device. Training loss Vs. validation loss and training accuracy Vs. validation accuracy charts are shown in Figure \ref{fig:gender-chart}.

\smallskip\noindent\textbf{CNN with Spectrogram}: We generate spectrograms for all word regions our algorithm has detected. We train our developed CNN model with the extracted spectrograms. We observed, at best 79.72\% accuracy on analyzing, which is also lower compared to what we get from classical machine learning algorithms. We can see the loss and accuracy graph in Figure \ref{fig:gender-acc-spec} and Figure \ref{fig:gender-loss-spec}. Details results on gender detection are listed in Table \ref{Gender}.

\begin{table}[t]
\scriptsize
\centering
\caption{Gender recognition accuracy (random guess 50\%).}
%\vspace{-3mm}
\begin{tabular}{|P{1.2cm}|P{1.4cm}|P{1.1cm}|P{1.5cm}|P{1.5cm}|}
 \hline
 \textbf{Method}& \textbf{Classifier} & \textbf{Data set} &\textbf{Accuracy (OnePlus 7T)}&\textbf{Accuracy (OnePlus 9)}\\
 \hline
 \hline
 \multirow{8}{6em}{Time and Frequency Domain Features} & \multirow{2}{6em}{Random Forest} & Emo-DB & \textbf{98.66\%}   &  88.67\%  \\
 \cline{3-5}
 & & JL-Corpus & 78.62\% &  77.71\%  \\
  \cline{2-5}
 &\multirow{2}{6em}{Random Subspace} & Emo-DB & 98.66\%  & 84.67\%   \\
 \cline{3-5}
 & & JL-Corpus & 79.37\% & 74.20\%  \\
\cline{2-5}
 &\multirow{2}{8em}{Decision Table} & Emo-DB & 98.21\%  & 84.67\%  \\
 \cline{3-5}
 & & JL-Corpus & 77.67\%  &  72.14\% \\
\cline{2-5}
  &\multirow{2}{8em}{CNN} & Emo-DB & 95.55\%  & 83.33\%  \\
  \cline{3-5}
 & & JL-Corpus & 75.17\%  & 67.52\%  \\
 \hline
 \multirow{2}{8em}{Spectrogram} & \multirow{2}{8em}{CNN}  & Emo-DB & 79.72\%  &  69.69\% \\
\cline{3-5}
 & & JL-Corpus & 70.10\% &  65.53\% \\
 \hline
 \end{tabular}
\label{Gender}
      %\vspace{-3mm}

\end{table}

\subsection{Speaker Detection}

We use JL-Corpus \cite{jlcorpus} and FSDD (Free Spoken Digit Dataset) \cite{FSDD} dataset to evaluate if the speaker's identity can be detected from the accelerometer data. We use the same ML algorithms and CNN classifiers that we have used in gender detection. 

\begin{figure}[t]
%\vspace{-7mm}
 \subfloat[Speaker recognition training loss Vs. validation loss for FSDD dataset (time-frequency feature analysis). \label{fig:speaker-loss}]{%
 \includegraphics[width=0.22\textwidth, height=3cm]{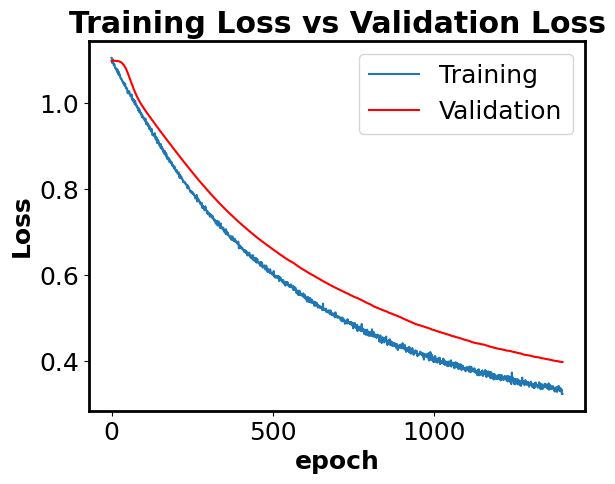}
 }
 \hfill
 \subfloat[Speaker recognition training accuracy Vs. validation accuracy for FSDD dataset (time-frequency feature analysis).\label{fig:speaker-acc}]{%
 \includegraphics[width=0.22\textwidth, height=3cm]{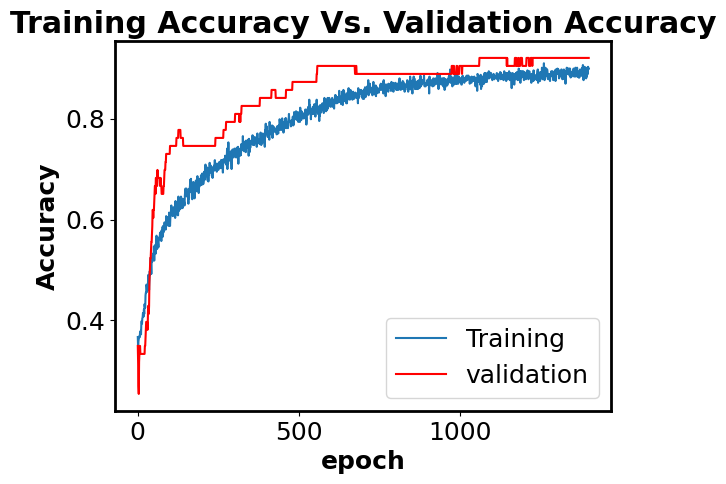}
 }
 \vfill

 \subfloat[Speaker recognition training loss Vs. validation loss for JL-Corpus dataset (spectrogram analysis). \label{fig:speaker-loss-spec}]{%
 \includegraphics[width=0.22\textwidth, height=3cm]{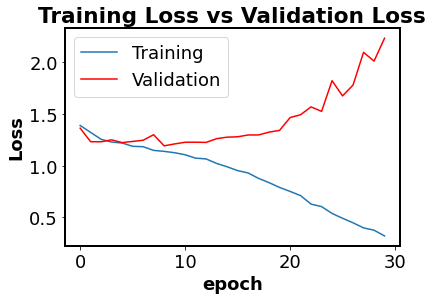}
 }
 \hfill
 \subfloat[Speaker recognition training accuracy Vs. validation accuracy for JL-Corpus dataset (spectrogram analysis). \label{fig:speaker-acc-spec}]{%
 \includegraphics[width=0.22\textwidth, height=3cm]{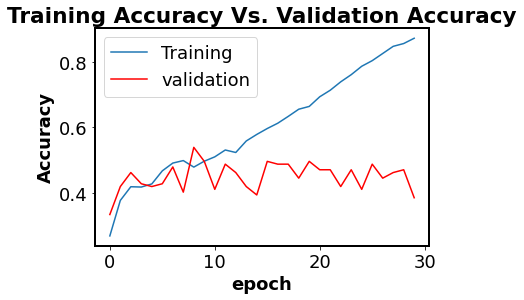}
 }
 	%\vspace{-2mm}
 \caption{Speaker recognition training and validation accuracy graph using different methods.}
\label{fig:speaker-chart}
  \vspace{-3mm}
 	
 \end{figure}

\smallskip\noindent\textbf{ML Algorithm with Time/Frequency Domain Features}: For the FSDD dataset, our detection program can detect 788 word regions among 1500 (1500 data was removed for inconsistent volume and background noise as discussed earlier) original utterances with three different classes for OnePlus 7T. For the OnePlus 9, the total number of extraction is 618. We extract all detected word regions' time/frequency domain features and use RandomForest, RandomSubspace, and Decision Tree as classifiers for our analysis. We use 80/20 train/test split and 10-Fold cross-validation similar to gender detection.

For RandomForest Classifier and FSDD dataset, we have achieved 91.24\% accuracy in classifying genders for the OnePlus 7T device and 87.75\% for the OnePlus 9. Whereas for RandomSubspace, we have also achieved 90.98\% and 88.70\% accuracy for these devices. For Decision Table, we have observed 90.22\% accuracy for the OnePlus 7T and 88.23\% accuracy for the OnePlus 9. The detailed result is shown in Table \ref{speaker}, Table \ref{result-op7}, and Table \ref{result-op9}.

Like previous analysis, for the JL-Corpus dataset, our detection program can detect 1469 word regions among 2400 utterances with four different classes. Here, we also have used the same classifiers to evaluate our results.

For RandomForest Classifier, we have achieved 64.60\% and 61.50\%  accuracy in classifying genders for OnePlus 7T and OnePlus 9. Whereas for RandomSubspace, we have also achieved 64.32\% and 59.86\% accuracy, and for Decision Table, we have observed 63.03\% and 55.72\% accuracy for the OnePlus 7T and OnePlus 9 devices, respectively. The detailed result is shown in Table \ref{speaker}, Table \ref{result-op7}, and Table \ref{result-op9}.

\smallskip\noindent\textbf{CNN with Time/Frequency Domain Features}: As discussed, we have extracted time and frequency domain features for 788 word regions with three different speakers for FSDD and 1469 word regions with four different speakers for JL-Corpus datasets. Our designed CNN is the same that we have used in gender detection.

We have used \textit{categorical\_crossentroy} as the loss function and \textit{Root Mean Square Propagation (RMSProp)} as the optimizer and 80/20 split as the train/test split in our analysis. We have achieved 86.07\% and 78.12\% for the FSDD dataset and 60.20\% and 57.73\% accuracy for the JL-Corpus dataset for OnePlus 7T and OnePlus 9 devices. Training loss Vs. validation loss and training accuracy Vs. validation accuracy charts are shown in Figure \ref{fig:speaker-chart}.

\smallskip\noindent\textbf{CNN with Spectrogram}: We generate spectrograms for all identified word regions and label them accordingly. After that, we fed generated spectrogram to our designed CNN-based image classifier. We collect data from two datasets (FSDD and JL-Corpus), and the CNN-based image classifier shows up to 45.23\% accuracy for the JL-Corpus dataset. We observe that accuracy is much lower than in classical machine learning algorithms. Loss and accuracy analysis are illustrated in Figure \ref{fig:speaker-acc-spec} and Figure \ref{fig:speaker-loss-spec}.

Details results on gender detection are listed in Table \ref{speaker}.

\begin{table}[t]
\centering
\scriptsize
\caption{Speaker recognition accuracy.}
%\vspace{-3mm}
\begin{tabular}{|P{1.2cm}|P{1.4cm}|P{1.1cm}|P{1.5cm}|P{1.5cm}|}
 \hline
 \textbf{Method}& \textbf{Classifier} & \textbf{Data set} &\textbf{Accuracy (Random Guess) (OnePlus 7T)}&\textbf{Accuracy (Random Guess) (OnePlus 9)}\\
 \hline
 \hline
 \multirow{8}{6em}{Time and Frequency Domain Features} & \multirow{2}{6em}{Random Forest} & FSDD & \textbf{91.24\%} (33\%) &  87.75\% (33\%)\\
 \cline{3-5}
 & & JL-Corpus & 64.60\% (25\%)&  61.50\% (25\%)\\
  \cline{2-5}
 &\multirow{2}{6em}{Random Subspace} & FSDD & 90.98\% (33\%) & 88.70\% (33\%)  \\
 \cline{3-5}
 & & JL-Corpus & 64.32\% (25\%)&  59.86\% (25\%)\\
\cline{2-5}
 &\multirow{2}{8em}{Decision Table} & FSDD & 90.22\% (33\%) & 88.23\% (33\%) \\
 \cline{3-5}
 & & JL-Corpus & 63.03\% (25\%) & 55.72\% (25\%) \\
\cline{2-5}
  &\multirow{2}{8em}{CNN} & FSDD & 86.07\% (33\%) & 78.12\% (33\%) \\
  \cline{3-5}
 & & JL-Corpus & 60.20\% (25\%) &  57.73\% (25\%)\\
 \hline
 \multirow{2}{6em}{Spectrogram} & \multirow{2}{8em}{CNN}  & FSDD &  35\% (33\%) & 36.44\% (33\%) \\
\cline{3-5}
 & & JL-Corpus & 44.32\% (25\%) & 45.23\% (25\%) \\
 \hline
 \end{tabular}
\label{speaker}
      %\vspace{-3mm}

\end{table}

\subsection{Speech Recognition}

As a representative speech recognition dataset, we used the FSDD (Free Spoken Digit Dataset) \cite{FSDD} dataset containing audio records of three different actors uttering digits 0 (zero) to 9 (nine). We evaluate the impacts on the accelerometer and try to find out if every digit can be distinguished using raw accelerometer data. We have used the same ML algorithms and CNN classifiers that we have used in gender detection. 

\smallskip\noindent\textbf{ML Algorithm with Time/Frequency Domain Features}: For the FSDD dataset, our detection program can detect 788 (OnePlus 7T)and 630 (OnePlus 9) word regions among 1500 (1500 data was removed for inconsistent volume and background noise as discussed earlier) original utterances with ten different classes. We have extracted all detected word regions' time/frequency domain features. We have used RandomForest, RandomSubspace, and Decision Tree as classifiers for our analysis, similar to gender and speaker analysis. We also use 80/20 train/test split and 10-Fold cross-validation similar to the previous analysis.

For RandomForest Classifier, we have achieved 54.46\% accuracy in classifying genders. Whereas for RandomSubspace, we have also achieved 54.46\% accuracy, and for Decision Table, we have observed 45.53\% accuracy for the OnePlus 7T. 

\smallskip\noindent\textbf{CNN with Time/Frequency Domain Features}: As discussed, we have extracted time and frequency domain features for word regions extracted for both phones with ten different speakers for FSDD. Our designed CNN is the same that we have used in gender and speaker detection.

\begin{figure}[t]
%\vspace{-7mm}
 \subfloat[Speech recognition training loss Vs. validation loss for FSDD dataset (time-frequency feature analysis). \label{fig:speech-loss}]{%
 \includegraphics[width=0.22\textwidth, height=3cm]{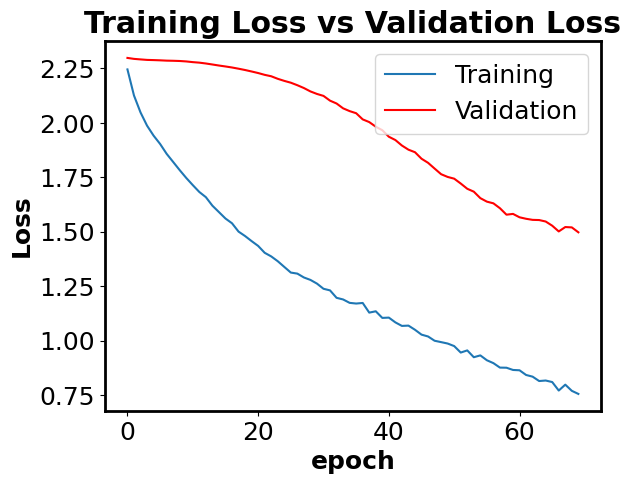}
 }
 \hfill
 \subfloat[Speech recognition training accuracy Vs. validation accuracy for FSDD dataset (time-frequency feature analysis). \label{fig:speech-acc}]{%
 \includegraphics[width=0.22\textwidth, height=3cm]{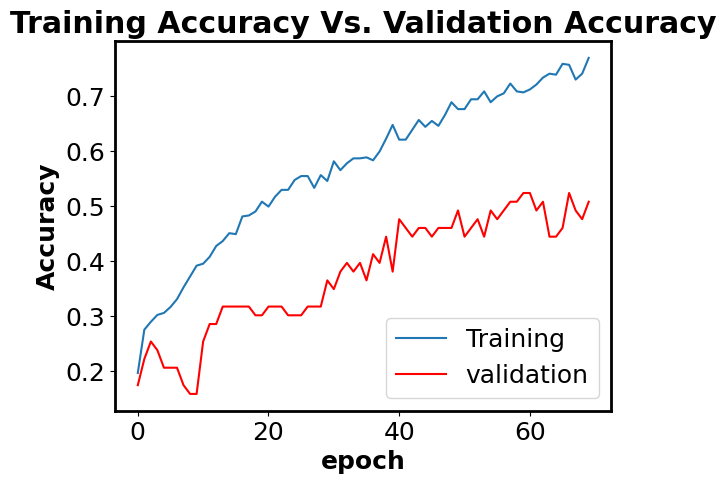}
 }

 	%\vspace{-2mm}
 \caption{Speech recognition training and validation accuracy graph using CNN with time-frequency features.}
\label{fig:speech-chart}
  \vspace{-3mm}
 	
 \end{figure}

Similar to speaker analysis, we have used \textit{categorical\_crossentroy} as the loss function and \textit{Root Mean Square Propagation (RMSProp)} as the optimizer and 80/20 split as the train/test split in our analysis. We have achieved 41.02\% and 38.70\% accuracy for the CNN analysis for OnePlus 7T and OnePlus 9 devices. Training loss Vs. validation loss and training accuracy Vs. validation accuracy charts are shown in Figure \ref{fig:speech-chart}.

The detailed result is shown in Table \ref{speech}, Table \ref{result-op7}, and Table \ref{result-op9}

\begin{table}[t]
\scriptsize
\centering
\caption{Speech recognition accuracy (random guess 10\%).}
%\vspace{-3mm}
\begin{tabular}{|P{1.2cm}|P{1.7cm}|P{0.9cm}|P{1.4cm}|P{1.4cm}|}
 \hline
 \textbf{Method}& \textbf{Classifier} & \textbf{Data set} &\textbf{Accuracy (OnePlus 7T)}&\textbf{Accuracy (OnePlus 9)}\\
 \hline
 \hline
 \multirow{4}{6em}{Time and Frequency Domain Features} & \multirow{1}{6em}{Random Forest} & FSDD & 53.59\%  & 41.59\%  \\
  \cline{2-5}
 &\multirow{1}{10em}{Random Subspace} & FSDD & \textbf{56.42}\%  &  38.99\% \\
\cline{2-5}
 &\multirow{1}{8em}{Decision Table} & FSDD & 51.80\%  &  33.33\% \\
\cline{2-5}
  &\multirow{1}{8em}{CNN} & FSDD & 41.02\%  & 38.70\% \\
 \hline
 \end{tabular}
\label{speech}
      %\vspace{-3mm}

\end{table}

\subsection{Result Summary}

\begin{table}[!hbtp]
\centering
\scriptsize
\caption{Detection performance of ML algorithm with time/frequency domain features for OnePlus 7T device.}
%\vspace{-3mm}
\begin{tabular}{|P{0.9cm}|P{1.0cm}|P{1.1cm}|P{0.7cm}|P{0.7cm}|P{0.9cm}|P{0.7cm}|}
 \hline
 \textbf{Detection}& \textbf{Classifier} & \textbf{Data set} &\textbf{TP Rate}&\textbf{FP Rate}& \textbf{Precision}& \textbf{Recall}\\
 \hline
 \hline
 \multirow{6}{6em}{Gender} & \multirow{2}{6em}{Random \\ Forest} & emo-DB & 98.7\% &  1.3\%  & 98.7\% & 98.7\%\\
 \cline{3-7}
 & & JL-Corpus & 78.6\% & 21.7\% & 78.8\% & 78.6\% \\
  \cline{2-7}
 &\multirow{2}{6em}{Random \\ Subspace} & emo-DB & 98.7\% & 1.3\% & 98.7\% & 98.7\%\\
 \cline{3-7}
 & & JL-Corpus & 79.4\% & 21.0\% & 79.8\% & 79.4\% \\
\cline{2-7}
 &\multirow{2}{8em}{Decision \\ Table} & emo-DB & 98.2\% & 1.9\% & 98.2\% & 98.2\%  \\
 \cline{3-7}
 & & JL-Corpus & 77.7\%  & 22.5\% & 77.7\% & 77.7\%  \\
 \hline
 \hline
 \multirow{6}{6em}{Speaker} & \multirow{2}{6em}{Random \\ Forest} & FSDD & 91.2\% & 4.6\% & 91.4\%  & 91.2\%  \\
 \cline{3-7}
 & & JL-Corpus & 64.6\% & 11.6\%  & 66.3\% & 64.6\% \\
  \cline{2-7}
 &\multirow{2}{6em}{Random \\ Subspace} & FSDD & 91.0\% & 4.7\% & 91.5\% & 91.0\% \\
 \cline{3-7}
 & & JL-Corpus & 64.3\% & 11.5\% & 67.5\% & 64.3\%  \\
\cline{2-7}
 &\multirow{2}{8em}{Decision \\ Table} & FSDD & 90.2\% & 5.1\% & 90.4\% & 90.2\% \\
 \cline{3-7}
 & & JL-Corpus & 63.0\%  & 11.9\% & 66.9\%  & 63.0\%  \\
 \hline
 \hline
 \multirow{3}{6em}{Speech} & Random Forest & FSDD & 53.6\% & 5.1\%  & 52.2\% & 53.6\% \\
  \cline{2-7}
 & Random Subspace & FSDD & 56.4\% & 4.8\% & 55.3\% & 56.4\%  \\
\cline{2-7}
 & Decision Table & FSDD & 51.8\% & 5.4\% & 50.9\% & 51.8\% \\
 \hline
 \end{tabular}
\label{result-op7}
      %\vspace{-3mm}

\end{table}

\smallskip\noindent\textbf{Gender Detection}: After evaluation with two different datasets and three different methods, we found reasonable accuracy with the highest achieved accuracy of 98.6\% in classical ML algorithm analysis for emo-DB datasets, which contains 10 different actors (5 male actors, 5 female actors). We have also evaluated with JL-Corpus dataset that contains 2400 utterances (we extracted 1469 word regions). JL-Corpus utterances are collected from 4 speakers (2 males and 2 females). We have achieved 79.37\% accuracy on the JL-Corpus dataset using the classical ML algorithm in gender detection, which is also a reasonable accuracy. CNN analysis with time and frequency domain features is also in line with the accuracy we get using ML algorithms (95.55\% accuracy with the FSDD dataset and 75.17\%).

Spearphone \cite{anand2021spearphone} shows the highest 99\% accuracy in gender detection using the motion sensor and loudspeaker of the smartphone. Considering the fact that ear speakers induced much lower vibration on the motion sensor, the gender recognition accuracy is almost similar to loudspeakers, which is an interesting observation. Another work Face-mic \cite{shi2021face}, use face vibrations on AR/VR devices and achieves 96.81\% accuracy in gender recognition. Compared to this work, we can say that vibration from ear speakers performs better in recognizing gender compared to face movements induced vibration.

\smallskip\noindent\textbf{Speaker Detection}: We evaluated with two different datasets (FSDD and JL-Corpus dataset). Among them, the FSDD datasets we used contains 788 utterances with 3 different actors and show 91.24\% accuracy using ML classifiers. The JL-Corpus dataset has 1469 utterances with four actors and shows the highest 64.60\% accuracy in speaker detection, which is still two times greater than a random guess. CNN analysis with time and frequency domain features also shows similar accuracy here (86.07\% for the FSDD dataset and 60.20\% for the JL-Corpus dataset).

Using the same handheld scenario, spearphone \cite{anand2021spearphone} achieved 99\% accuracy for one device in classifying ten speakers. On the other hand, we achieve, at best 91.24\% accuracy in classifying three speakers. Although loudspeaker performance is better than ear speakers in this case, the observed accuracy reveals the potential vulnerability of identifying speaker-specific information from ear speakers just using built-in zero permission motion sensors. It also shows slightly lower but still good accuracy compared to face movement-induced vibration (Face-Mic \cite{shi2021face}).

\smallskip\noindent\textbf{Speech Detection}: To evaluate speech detection, we have used the digit dataset FSDD, where three actors utter ten different digits. We evaluate the time and frequency domain features with classical ML algorithms, which show the highest 56.42\% accuracy. As there are ten different classes here, the accuracy still exhibits five times greater accuracy than a random guess, which implies that vibration due to the ear speaker induced a reasonable amount of distinguishable impact on accelerometer data.

Previous works also show lower accuracy in speech detection compared to gender and speaker detection. Spearphone \cite{anand2021spearphone}. shows 80\% accuracy in recognizing digits. In work, mmSpy \cite{basak2022mmspy}, which uses an external receiver to sense vibration from the ear speaker, can achieve 83\% accuracy at 1 ft distance and 47.99\% accuracy at 6 ft distance. Compared to a practical attack scenario (4-6 ft of distance from the phone), our result is promising in detecting speech digit data.

\smallskip\noindent\textbf{Other Performance Evaluation Metrics}: We list down TP-rate, FP-rate, Precision, and Recall of our analysis using classical machine learning algorithm in Table \ref{result-op7} and Table \ref{result-op9}. 

TP Rate (True Positive Rate) indicates the rate of correctly classified elements. FP Rate (False Positive Rate) shows the rate of incorrectly classified elements for a particular class. FP Rate (False Positive Rate) shows the rate of incorrectly classified elements for a particular class. Precision indicates the proportion of correctly classified elements and all classified elements for a particular class. On the other hand, recall suggests the proportion of correctly classified elements and all the elements present in the class.

\begin{table}[t]
\centering
\scriptsize
\caption{Detection performance of ML algorithm with time/frequency domain features for OnePlus 9 device.}
%\vspace{-3mm}
\begin{tabular}{|P{0.9cm}|P{1.0cm}|P{1.1cm}|P{0.7cm}|P{0.7cm}|P{0.9cm}|P{0.7cm}|}
 \hline
 \textbf{Detection}& \textbf{Classifier} & \textbf{Data set} &\textbf{TP Rate}&\textbf{FP Rate}& \textbf{Precision}& \textbf{Recall}\\
 \hline
 \hline
 \multirow{6}{6em}{Gender} & \multirow{2}{6em}{Random \\ Forest} & emo-DB & 88.7\%  &  11.8\%  & 88.7\% & 88.7\%\\
 \cline{3-7}
 & & JL-Corpus & 78.6\% & 21.7\% & 78.8\% & 78.6\% \\
  \cline{2-7}
 &\multirow{2}{6em}{Random \\ Subspace} & emo-DB & 84.7\% & 15.4\% & 84.7\% & 84.7\%\\
 \cline{3-7}
 & & JL-Corpus & 79.4\% & 21.0\% & 79.8\% & 79.4\% \\
\cline{2-7}
 &\multirow{2}{8em}{Decision \\ Table} & emo-DB & 84.7\% & 16.7\% & 84.8\% &  84.7\% \\
 \cline{3-7}
 & & JL-Corpus & 77.7\%  & 22.5\% & 77.7\% & 77.7\%  \\
 \hline
 \hline
 \multirow{6}{6em}{Speaker} & \multirow{2}{6em}{Random \\ Forest} & FSDD & 87.8\% & 5.8\% & 87.9\%  & 87.8\%  \\
 \cline{3-7}
 & & JL-Corpus & 61.5\% & 13.2\%  & 61.1\% & 61.5\% \\
  \cline{2-7}
 &\multirow{2}{6em}{Random \\ Subspace} & FSDD & 88.7\% & 5.2\% & 89.1\% & 88.7\% \\
 \cline{3-7}
 & & JL-Corpus & 55.7\%  & 15.5\% & 55.5\% &  55.7\% \\
\cline{2-7}
 &\multirow{2}{8em}{Decision \\ Table} & FSDD & 88.2\% & 5.4\% & 88.8\% & 88.2\% \\
 \cline{3-7}
 & & JL-Corpus & 59.9\%  & 13.7\% & 59.6\%  & 59.9\%  \\
 \hline
 \hline
 \multirow{3}{6em}{Speech} & Random Forest & FSDD & 41.6\% & 6.8\%  & 41.6\% & 41.6\% \\
  \cline{2-7}
 & Random Subspace & FSDD & 39.0\% & 7.2\% & 39.1\% & 39.0\%  \\
\cline{2-7}
 & Decision Table & FSDD & 33.3\% & 8.0\% & 33.6\% & 33.3\% \\
 \hline
 \end{tabular}
\label{result-op9}
      %\vspace{-3mm}

\end{table}

%% file: discussion.tex
\subsection{General Discussion}

\smallskip\noindent\textbf{Sensor Rate Limit}: To protect potentially sensitive information about users, if the app targets Android 12 (API level 31) or higher, the system has a limit on the refresh rate of data from certain motion sensors and position sensors. This data includes values recorded by the device's accelerometer. However, we performed gender classification by utilizing the emo-DB dataset and accomplished 90.97\% accuracy at a 200 Hz sampling rate, which is still a high reasonable accuracy. In this case, the restriction for the sensor rate does not impact much on the eavesdropping threat.

\smallskip\noindent\textbf{Band-pass Filter for Human Movement}: Currently, we are collecting all the data in one go. However, if we place a high-pass filter, the filter would still severely wipe out speech properties. For the ear speaker case, even if we placed a 1 Hz high pass filter, it reduces all speech properties. We conducted another test where we checked the accuracy when the phone was placed on a phone holder clamp, and no movement like the human hand gesture was involved but held in a handheld position.  We achieved 97.3\% for gender classification accuracy with the emo-DB dataset in the clamp testing. The results imply that human movement does not play any significant role in that. Thus, we do not remove any speech property by placing a high pass filter.

\subsection{Limitation}

Although recent smartphones use larger and more powerful ear speakers, they still reduce the volume at a reasonable level to ensure the comfort of the users during a phone conversation. As a result, they cannot generate a significant impact on raw accelerometer data. For this reason, our word region detection algorithm cannot detect a high percentage of the word uttered (it can detect 45\% - 80\% of words or speech in total). However, our result indicates that it is sufficient for the adversary to reasonably detect significant speech features (e.g., gender, speaker's identity, speech).

Impacts on the accelerometer due to the vibration of motion sensors are highly dependent upon the distance between the ear speaker and the motion sensor. It depends on the design of the smartphone's motherboard, which varies from manufacturer to manufacturer and even varies from model to model of the same manufacturer. So, our observed accuracy is not constant and can show slightly different results in terms of accuracy.

The data collected from the accelerometer can be noisy due to the hand and body movement of the user. However, as discussed in the previous subsection, even if we place 1 Hz high pass filter, it removes important speech features due to a very low impact on accelerometers by ear speakers. As such, removing low-frequency noises, in this case, is a challenge. However, we conducted another experiment where we emulated the handheld scenario with the phone attached to a clamp where there was no body-induced vibration present. We got 97.3\% accuracy compared to 98.6\% accuracy in our result, which implies there is a very low impact of noises in determining the accuracy.

\subsection{Countermeasures}

One of the potential countermeasures is to change the permission model of motion sensors so that third-party apps cannot record sensor data without the permission of users. Recently, Android has restricted sensor data collection without
permission \cite{googlerestriction200} for sampling rates beyond 200 Hz. However, this does not completely prevent silent eavesdropping using motion sensors. As discussed in the previous subsection, we have done another experiment on Gender detection where we collected all data at 200 Hz sampling rate instead of the default sampling rate (e.g., 420 Hz for OnePlus 7T and 520 Hz for OnePlus 9). We got 90.97\% accuracy in gender detection compared to 98.6\% accuracy we observed with the default sampling rate for OnePlus 7T phone. As such, this countermeasure cannot fully prevent the user from silent eavesdropping. 

Smartphone manufacturers should be more careful about designing larger and more powerful ear speaker volume control. They should maintain the same sound pressure during phone conversation as previous generation phones ear speakers. Moreover, they should place the motion sensors in the proper position relative to the ear speaker so that the phone speaker’s vibration impact can be minimized.

\subsection{Future Works}

Our work opens immense research opportunities on eavesdropping possibilities on ear speakers using smartphones' built-in sensors. As the ear speaker has very little impact on accelerometer data, it is always a challenge to effectively extract all word regions from it. Researchers can solve the challenge by proposing more efficient algorithms that increase detection rates. They also have the opportunity to design machine learning or deep learning techniques to achieve more accuracy in speech information extraction in general. 

This research primarily focuses on speech recognition and some speech feature (e.g., speaker's identity, gender) eavesdropping from the ear speaker-induced vibration in an accelerometer. Researchers can work on other speech features (e.g., language) extraction from motion sensor data. They also have the opportunity to work on potential preventive and mitigating measures for eavesdropping from ear speakers.

%% file: conclusion.tex
This work focused on the unexplored area of eavesdropping possibility using smartphone ear speakers, especially with the device equipped with multiple powerful speakers that are used as ear speakers. We investigate the reverberation effect of ear speakers on a built-in accelerometer by extracting time-frequency domain features and spectrograms. We evaluate them using classical machine learning algorithms and our developed convolutional neural network (CNN) models. We found up to 98.6\% accuracy on gender detection, up to 92.6\% accuracy on speaker detection, and up to 56.42\% accuracy on speech detection, which proves the presence of distinguishing speech features in the accelerometer data that the adversaries can leverage for eavesdropping. Our findings also open the opportunity for researchers to explore recently popular powerful ear speakers' potential risk factors.